\documentclass{article}

\usepackage{arxiv}

\usepackage[utf8]{inputenc} 
\usepackage[T1]{fontenc}    
\usepackage{url}            
\usepackage{booktabs}       
\usepackage{amsfonts}       
\usepackage{amsmath, amssymb}
\usepackage{nicefrac}       
\usepackage{microtype}      
\usepackage{lipsum}		
\usepackage{graphicx}
\usepackage[numbers]{natbib}
\usepackage{natbib}
\usepackage{doi}
\usepackage{xcolor}
\usepackage{enumitem}
\usepackage{bm}
\usepackage{multirow}
\usepackage{placeins}
\usepackage{todonotes}
\usepackage[ruled, linesnumbered, nofillcomment]{algorithm2e}
\usepackage{float}

\usepackage{color}
\usepackage{hyperref}       

\SetCommentSty{mycommfont}

\newcommand\changes[1]{\textcolor{black}{#1}} 
\newcommand\newchanges[1]{\textcolor{black}{#1}}

\title{Towards Understanding Structure-Function Relationships in Random Fiber Networks}

\date{\today}	

\author{\href{https://orcid.org/0009-0000-3325-1410}{\includegraphics[scale=0.06]{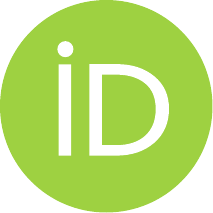}\hspace{1mm}Peerasait~Prachaseree}\\
	Department of Mechanical Engineering\\
	Boston University\\
	Boston, MA 02215 \\
	\texttt{pprachas@bu.edu} \\
	\And
	\href{https://orcid.org/0000-0001-8099-3468}{\includegraphics[scale=0.06]{orcid.pdf}\hspace{1mm}Emma~Lejeune} \\
	Department of Mechanical Engineering\\
	Boston University\\
	Boston, MA 02215\\
	\texttt{elejeune@bu.edu} \\
}

\renewcommand{\shorttitle}{Structure-Function Relationships of Fiber Network}

\hypersetup{
pdftitle={Structure-Function Relationships of Fiber Network},
pdfsubject={
    cond-mat.dis-nn, cond-mat.soft, physics.bio-ph},
pdfauthor={Peerasait~Prachaseree,Emma~Lejeune},
pdfkeywords={random fiber network, mesoscale, structure-function, structure-property, force chains, strain-stiffening, fiber recruitment},
}

\begin{document}
\maketitle

\begin{abstract}

Random fiber networks form the structural foundation of numerous biological tissues and engineered materials. From a mechanics perspective, understanding the structure-function relationships of random fiber networks is particularly interesting because when external force is applied to these networks, only a small subset of fibers will actually carry the majority of the load. Specifically, these load-bearing fibers propagate through the network to form load paths, also called force chains. However, the relationship between fiber network geometric structure, force chains, and the overall mechanical behavior of random fiber network structures remains poorly understood. To this end, we implement a finite element model of random fiber networks with geometrically exact beam elements, and use this model to explore random fiber network mechanical behavior. Our focus is twofold. First, we explore the mechanical behavior of single fiber chains and random fiber networks. Second, we propose and validate an interpretable analytical approach to predicting fiber network mechanics from structural information alone. \changes{Key findings include insight into the critical strain-stiffening transition point for single fiber chains and fiber networks generated from a Voronoi diagram, and a connection between force chains and the distance-weighted graph shortest paths that arise by treating fiber networks as spatial graph structures.} This work marks an important step towards mapping the structure-function relationships of random fiber networks undergoing large deformations. Additionally, with our code distributed under open-source licenses, we hope that future researchers can directly build on our work to address related problems beyond the scope defined here.
 
\end{abstract}
  
\section{Introduction}
\label{sec:intro}

\begin{figure}[h]
    \centering
    \includegraphics[width= 0.75\textwidth]{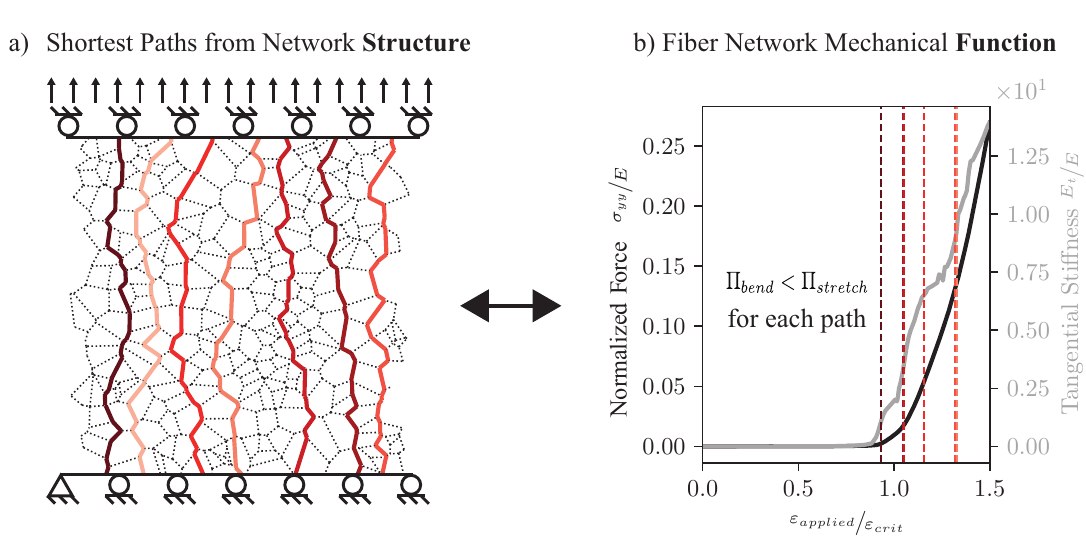}
    \caption{Graphical summary of our work. a) Schematic of distance-weighted shortest paths of our random fiber networks. b) Schematic demonstrating the role of shortest paths in the fiber-recruitment effect.}
    \label{fig:intro}
\end{figure}

Fiber network structures are ubiquitous in both natural and engineered systems. For example, in biological systems, fiber networks underlie the structure of soft tissues as diverse as the heart \citep{kakaletsis2021right}, the skin \citep{lin2024sex}, and the cornea~\citep{benoit2016simultaneous}. Random fiber networks also play an important role in many biological processes such as cell-cell communication \citep{humphries2017mechanical, mann2019force}, tissue contraction \citep{depalma2024matrix}, and wound healing \citep{das2021extracellular}. And, a myriad of common materials, such as paper \citep{ostoja2000random} and fabric \citep{ding2024unravelling}, are structurally composed of dense networks of fibers. More recently, there has also been interest in expanding the design space of architected materials towards bio-inspired random fiber networks structures \citep{reid2018auxetic}.

Across biological and engineered systems, random fiber networks exhibit interesting emergent mechanical behavior stemming from their complex structure. Understanding structure-function relationships in random fiber networks is relevant both to gaining a fundamental understanding of biological systems and biomechanical behavior, and to advancing materials design. 
However, analyzing the mechanical behavior of random fiber networks is non-trivial -- particularly when it comes to assessing behavior across length scales \citep{hatami2009heterogeneous,head2003distinct, picu2011mechanics}, and in the large deformation regime \citep{feng2016nonlinear, glaesener2019continuum}. In this work, we are specifically interested in random fiber networks in the mesoscale range -- where there is no distinct length-scale separation between fiber lengths and domain size.  Due to this lack of clear scale separation, the structure-function relationships of mesoscale fiber networks are not amenable to straightforward homogenization techniques and multi-scale analysis \citep{gilchrist2014micro, jimenez2023multiscale}. However, understanding how forces propagate through fiber networks at this length scale is highly relevant both to understanding cell signaling \citep{beroz2017physical,berthier2024nonlinear} and to the design of next generation metamaterials \citep{liu2021design, pahlavani2024deep}. As a step towards establishing a structure-function relationships in these mesoscale networks in the large deformation regime, we investigate the role of emergent load paths in the strain-stiffening effect of random fiber networks, as well as their connection to the fiber network geometric structure.

Due to the heterogeneous structure of mesoscale random fiber networks, local strain fields are different from far-field applied strain \citep{hatami2009heterogeneous}. This leads to a wide variety of interesting emergent properties. In particular, when external force is applied to random fiber networks, only a small subset of fibers carry a majority of the load. These load-bearing fibers join to form load paths, also known as force chains. Force chains have been used to analyze the fracture behavior of polymers \citep{deng2023nonlocal}, long-range force transmission \citep{kumar2023range, ronceray2016fiber, wang2014long}, and the strain-stiffening effect \citep{parvez2023stiffening,sarkar2022evolution, vzagar2015two}. More broadly, these force chains are also observed in granular systems and have been used to analyze shear bands \citep{karapiperis2021nonlocality,tordesillas2010force} and sound propagation \citep{owens2011sound}. Fundamentally, force chains are responsible for the non-local behavior of both granular systems and random fiber networks. \changes{However, identifying these force chains for both granular systems and random fiber networks typically relies on force information, which can be difficult to retrieve in experimental setups. Furthermore, despite numerous links between the emergence of force chains to overall mechanical response, the specific link between force chains and the geometric structure is not well understood. To address this, we will attempt to relate force chains to the geometric structure of random fiber networks through their distance-weighted graph shortest paths. In addition to identifying force chains, we will also associate these predicted force chains with the global emergent mechanical response of random fiber networks.} 

Our goal is to conduct research in this niche and thus contribute to the broader understanding of the mechanics of random fiber networks. Broadly speaking, modeling the mechanical behavior of fiber networks is a formidable challenge in part because most fiber network based materials contain multiple relevant length scales \citep{picu2011mechanics}. One method to model these fiber networks is to directly represent the fibers discretely. This allows the microstructure of the fiber network to be represented exactly. Previously, there has been substantial work towards mapping structure-function/property relationships of discrete random fiber networks. For example, in the small-strain regime, random fiber networks can deform affinely or non-affinely based on network density \citep{head2003distinct}, connectivity \citep{wyart2008elasticity}, and orientation \citep{hatami2009effect}. These deformation modes are dependent on the dominating energy mode in the fiber network and determine the mechanical behavior of fiber networks. The fibers in networks that deforms non-affinely are dominated by bending energy and are ``floppy'', while the fibers in networks that deforms affinely are stretching dominated and hence ``rigid'' \citep{heussinger2006floppy,wyart2008elasticity}. In terms of mechanical behavior, ``floppy'' networks are significantly less stiff than ``rigid'' networks. Crucially, in the finite strain regime, fiber networks can transition between a floppy mode to a rigid mode \citep{arzash2020finite,feng2016nonlinear,sharma2016strain}, which manifests as a strain-stiffening effect. This strain-stiffening effect emerges from the complex  interactions between random fiber networks, but the influence of the fiber network geometric structure on the strain-stiffening effect remains an open question. In general, most of the previously mentioned studies employ global statistical descriptors such as mean connectivity, mean orientation, or density  to understand the mechanics of random fiber networks. These global descriptors effectively parametrizes the geometric structure of the fiber networks, but insight on the interactions between the local heterogeneous geometric features and the global emergent mechanical response is lost. To this end, in this work, we aim to connect specific local geometric features to the global mechanical response through the distance-weighted shortest paths through the network.

Notably, there is also a rich history of modeling random fiber networks as a continuum. When considering random fiber networks as a continuum, information about the microstructure can be integrated through modeling fiber dispersion~\citep{holzapfel2019fibre}, incorporating a fiber deformation model \citep{britt2022constitutive},  volume averaging with representative volume elements \citep{dirrenberger2014towards, hatami2009heterogeneous}, or mapping the microscopic response to the microscopic response with data-driven approaches \citep{leng2021predicting}. While these continuum scale models are able to model the mechanics of random fiber networks well if the affine assumption holds, non-affine behavior is still non-trivial to capture with classical continuum models \citep{picu2021constitutive,rastak2018non, tkachuk2012maximal}. And, dealing with structures in the mesoscale range, where there is no clear length scale separation, is not feasible with continuum models. Most crucially, these models are not able to explicitly study the local deformation behavior of random fiber networks and its connection to global emergent behavior \citep{stracuzzi2022risky}. As such, in this work, we focus exclusively on discrete fiber network modeling.

\changes{Here, we computationally model discrete random fiber networks as geometrically exact beams and perform finite element analysis (FEA) to examine the mechanics of random fiber networks that are generated from Voronoi diagrams.} As our starting point, we first investigate the mechanics of single fiber chains, which is a simpler system with mechanical behavior highly relevant to fiber networks. We then map the structure-function relationship of these single fiber chains by both predicting the mechanics and critical strain-stiffening transition point through a simple analytical reduced order model that contains the geometric and kinematic features of the chain. The information obtained in single fiber chains is then translated to random fiber networks by interpreting the network's distance-weighted graph shortest paths as single fiber chains in parallel, as illustrated in Fig \ref{fig:intro}a. \changes{Finally, the graph shortest paths are linked to the mechanical functions of random fiber networks, such as the emergence of force chains, and the strain-stiffening effect in random fiber network that stems from fiber recruitment, as shown in Fig. \ref{fig:intro}b.} Overall, this work is aimed as a step towards mapping the structure-function relationships of random fiber networks by linking complex local geometric structure to its global emergent mechanical behavior.


\newchanges{
Because the manuscript includes substantial background and methodological detail to support reproducibility, we provide the following guide to help readers quickly identify the key novel contributions and their locations within the text:
\begin{itemize}
    \item We introduce an open-source implementation of a computational model for random fiber networks using FEniCS in Section \ref{sec:meth_model}. This section begins with the general 2D beam formulation (Section \ref{sec:meth_2D}), followed by our approach to alleviating numerical locking  (Section \ref{sec:meth_hybrid}) and handling local instabilities (Section \ref{sec:meth_regularize}). 
    \item The proposed analytical reduced order model is presented in Section \ref{sec:meth_rom}. We begin by formulating the model for single fiber chains (Section \ref{sec:meth_rom_single})  and then extend it to full fiber networks using a distance-weighted graph shortest path approach (Section \ref{sec:meth_rom_network})
    \item We analyze the behavior of the reduced order model for single fiber chains in Section \ref{sec:res_analytical_single}, and in Section \ref{sec:res_single_phase}, we relate geometric features of individual fibers to strain stiffening behavior.
    \item In Section \ref{sec:res_fiber_abc}, we analyze the reduced order model applied to fiber networks. We then investigate the relationship between network stiffening and the distance-weighted graph shortest path in Sections \ref{sec:res_forcechain} and \ref{sec:res_recruitment}. Section \ref{sec:res_fiber_phase} outlines the limitations of this approach, and Section \ref{sec:res_ablation} presents a potential application.
\end{itemize}
}

\section{Methods}
\label{sec:meth}

In this work, we are interested in understanding the connection between the geometry and mechanics of random fiber networks through a computational model. To achieve this goal, we begin by understanding the relationship between single fiber chain geometry and single fiber chain mechanics, and then extrapolate these insights to random fiber networks. We first introduce our computational model as well as detail our approach to address challenges with numerical convergence in Section \ref{sec:meth_model}. Then, we propose a simple analytical reduced order model to connect the geometry and kinematics to the mechanics of both single fiber chains and fiber networks in Section \ref{sec:meth_rom}. Finally, we end in Section \ref{sec:meth_geo} by introducing the geometric domain and boundary conditions that we use for our computational investigation presented in Section 3.

\subsection{Computational Model}
\label{sec:meth_model}
In this Section, we introduce our finite element based computational pipeline to simulate both single fiber chains and random fiber networks. In brief, we model single fiber chains as beams elements within a finite element framework using the open source software platform FEniCS \citep{alnaes2015fenics,logg2012automated}. \changes{We introduce the 2D beam formulation used to model our random fiber networks in Section \ref{sec:meth_2D}}. To enable the numerical convergence of our finite element model, we use hybrid beam elements as described in Section \ref{sec:meth_hybrid}, and Tikhonov regularization as introduced in Section \ref{sec:meth_regularize}. Finally, in Section \ref{sec:meth_fea}, we introduce the finite element problem and perform a brief analysis of the discretized formulation.

\subsubsection{\changes{Geometrically Exact Beams in 2D}}
\label{sec:meth_2D}
\begin{figure}[ht]
    \centering
    \includegraphics[width= 0.90\textwidth]{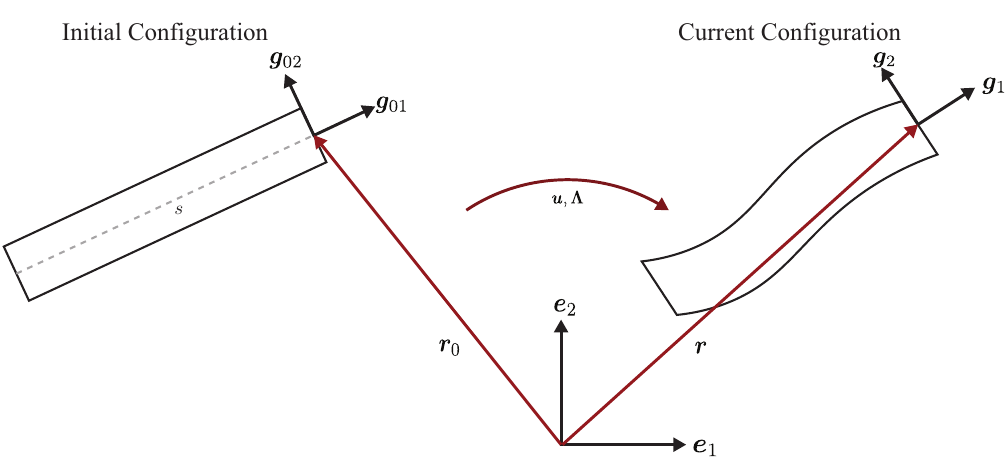}
    \caption{\changes{Schematic of the kinematics defining initial configuration and deformed configuration of 2D geometrically exact beams. The initial beam position $\bm{r}_0(s)$ is mapped to the current beam configuration $\bm{r}(s)$ through the displacement field $\bm{u}$, and the initial beam orientation $\{ \bm{g}_{01}, \bm{g}_{02} \}$ is mapped to the current beam orientation $\{ \bm{g}_1, \bm{g}_2 \}$ through the rotation field $\bm{\Lambda}$.}}
    \label{fig:meth_beams_2D}
\end{figure}

\changes{To capture the mechanical behavior of slender fibers in our fiber networks, we model each fiber as shear-deformable geometrically exact beams. These geometrically exact beams are also be referred to as Simo-Reissner beams and nonlinear spatial Timoshenko beams in the literature \citep{meier2019geometrically,reissner1981finite,simo1985finite}. In general, geometrically exact beams can be non-trivial to implement due to singularities associated with rotation vector parameterizations \citep{ibrahimbegovic1997choice,magisano2020large, meier2019geometrically}. For completeness, we outline the 3D geometrically exact beam formulation in Appendix \ref{appendix:meth_beams}. However, in modeling 2D fiber networks, we are able to make simplifications to the full 3D beam model. First, we assume that our fiber networks can only deform in-plane (i.e., in the $e_1$ and $e_2$ directions; see Fig. \ref{fig:meth_beams_2D}), and will not be subjected to external torsion and moment. Similar to the 3D case, we also describe the position of the beam centerline with a parameterized curve $s \in[0,\ell]$ where $\ell$ is the beam length, and the beam initial beam orientation is parameterized with a field of orthonormal bases $\{ \bm{g}_{1}(s), \bm{g}_{2}(s) \} \in \mathbb{R}^2$. The initial beam configuration with respect to the global coordinates is described by $\bm{r}_0(s) \in \mathbb{R}^2$. The initial beam configuration $\bm{r}(s) \in \mathbb{R}^2$ can be mapped to the current beam configuration through the displacement field $\bm{u}(s) \in \mathbb{R}^2$ such that:}
\changes{
\begin{equation}
    \bm{r}(s) = \bm{r}_0 + \bm{u}(s) \, .
\end{equation}
}
\changes{
Similarly, the initial orthonormal bases $\{ \bm{g}_{01}, \bm{g}_{02}\}$ can be mapped to the current bases $\{ \bm{g}_{1}, \bm{g}_{2}\}$ through a rotation tensor $\bm{\Lambda} \in SO(2)$ such that:}

\changes{
\begin{equation}
    \bm{g}_{i} = \bm{\Lambda} \bm{g}_{0i} \quad ; \quad i = 1,2 \, .
\end{equation}}

\changes{Typically, the direction of the first initial bases is tangent to the beam direction (i.e., $\bm{g_{01} = r_{0,s}}$, where $(.)_{,s} = \frac{d(.)}{ds}$). In the 2D case, the rotation tensor can be constructed simply through the scalar rotational degree of freedom $\theta(s) \in \mathbb{R}$ such that:}

\changes{
\begin{equation}
    \bm{\Lambda} = \begin{bmatrix} 
    \text{cos}\theta & -\text{sin}\theta \\  
    \text{sin}\theta & \text{cos}\theta 
    \end{bmatrix}
\end{equation} \, .
}

\changes{Next, the strain measures are defined as:}

\changes{\begin{equation}
    \begin{aligned}
        \bm{\varepsilon} &= \bm{\Lambda}^T\bm{r}_{,s} - \bm{g}_{01}\\
        \chi &= \theta_{,s} \,  
    \end{aligned}
\end{equation}
}
\changes{
where $\bm{\varepsilon}$ and $\chi$ are the translational and rotational strain respectively. Finally, the constitutive matrices proposed by Simo \citep{simo1985finite} can be simplified. The translational constitutive matrix can be simplified to $\bm{C}_N = \text{diag}[EA, \mu A^*]$ with $\text{diag}[\dots]$ denoting a diagonal matrix, $E$ and $\mu$ are the Young's Modulus and shear modulus respectively, $A$ and $A^*$ are the cross-section area and shear-corrected areas respectively. The rotational constitutive relation also simplifies to $C_M = EI$, with $I$ denoting the second moment of area. As such, the hyperelastic strain energy density can be written as:}
\changes{
\begin{equation}
   \tilde{\Pi}_{int} = \frac{1}{2} \bigg(  \bm{\varepsilon} \bm{\cdot} \bm{C}_N \bm{\varepsilon} + C_M \chi^2 \bigg)\, . 
   \label{eqn:beam_int}
\end{equation}}

\changes{
Then, by grouping the generalized strain terms $\bm{\tilde{\varepsilon}} = [\bm{\varepsilon}, \chi]$, and defining $\mathbb{C} = \text{diag}[\bm{C}_N,C_M]$, the total internal beam strain energy can be written as:}

\changes{
\begin{equation}
    \label{beam_final}
    \Pi_{int} = \int_\ell \tilde{\Pi}_{int} ds = \frac{1}{2} \int_{\ell} \left[ \bm{\tilde{\varepsilon}} \cdot \mathbb{C} \bm{\tilde{\varepsilon}} \right] ds \, .
\end{equation}}

\changes{
While the above beam formulation works for a myriad of structural mechanics problems, random fiber networks formulated with geometrically exact beam models still experience numerical convergence difficulties such as locking and local instabilities. The next two sections below detail our approach to address these numerical convergence issues.}

\subsubsection{Alleviating Locking with Consistent Interpolation and Hybrid Beam Elements}
\label{sec:meth_hybrid}

Since we model our single fiber chains and random fiber networks as slender beams using a shear deformable beam formulation, shear locking can be observed in the system if no steps are taken to address it. In general, shear locking occurs from the difficulty shear deformable beam elements have in satisfying the Kirchhoff constraint (i.e., vanishing shear strains in the limit of slender beams under pure bending) \citep{belytschko2014nonlinear}. To alleviate shear locking, we discretize the beams using mixed elements with quadratic Lagrange interpolation functions for displacements, and linear Lagrange interpolation functions for rotations. This consistent interpolation approach alleviates most shear locking issues in slender beams~\citep{reddy1997locking}. 

In addition to shear locking, slender beams undergoing finite rotation can also exhibit extrapolation locking behavior. In brief, extrapolation locking occurs when the tangential stiffness matrix is no longer a good approximation of the equilibrium curve \citep{garcea1998mixed, magisano2017advantages}. To alleviate extrapolation locking, we turn to hybrid beam elements. Hybrid beam elements are formulated through the Hellinger-Reissner Principle which formulates beam internal energy in terms of the complementary energy such that Eqn. \ref{beam_final} becomes:

\begin{equation}
    \Pi_{int} =  \int_\ell \left[ \bm{\tilde{\sigma}} \cdot \bm{\tilde{\varepsilon}} - \frac{1}{2}\left( \bm{\tilde{\sigma}} \cdot \mathbb{C}^{-1} \bm{\tilde{\sigma}} \right) \right] ds
\end{equation}

where $\bm{\tilde{\sigma}} = [\bm{\Lambda^T n}, \bm{m}]$ with $\bm{n}$ and $\bm{m}$ denoting the internal force and moment respectively \citep{reissner1950variational}. Note that the Hellinger-Reissner Principle converts the internal force/moment into an additional independent unknown and essentially changes the minimization problem to a saddle-point problem. The stationary conditions for the Hellinger-Reissner Principle are:

\begin{equation}
    \label{eqn:stationary}
    \begin{aligned}
             & \Pi_{int}' \delta \bm{\tilde{u}} =  \int_{\ell} \left[ \delta\bm{\tilde{\varepsilon}} \cdot \bm{\tilde{\sigma}} \; \right] ds = 0 \\
             & \Pi_{int}' \delta \bm{\tilde\sigma} = \int_{\ell} \left[ \delta \bm{\tilde{\sigma}} \cdot \big( \bm{\tilde{\varepsilon}} - \mathbb{C}^{-1} \bm{\tilde{\sigma}} \big) \right] \; ds = 0
        \end{aligned}
\end{equation}

where $(.)'$ denotes taking the first-order functional derivative. The first term obtained from taking the variation with respect to the displacements is the virtual work term and enforces equilibrium, while the second term obtained by taking the variation with respect to the stresses is the constitutive model which now acts as a weak constraint. As a result, it is not necessary that the constitutive relations are satisfied during the Newton iterations as the stresses are independent variables. However, the constitutive relations will be satisfied at equilibrium. This results in a smoother change of tangential stiffness matrix during the Newton iterations and enhances convergence \citep{garcea1998mixed, magisano2017advantages}. In this work, we use linear discontinuous (i.e., DG1) elements and constant discontinuous (i.e., DG0) elements to interpolate the force and moment respectively. Note that the displacement degrees of freedom still use Lagrange quadratic interpolation functions and the rotation degrees of freedom use linear Lagrange interpolation functions. This choice of interpolation functions have been shown to work well in a slender beam problems, and in certain cases, are equivalent to reduced integration of beam elements, which has the also further reduces the effect of shear locking \citep{noor1981mixed}. A more detailed analysis on the hybrid elements used in this work can be found in Section \ref{sec:meth_fea}.

\subsubsection{Addressing Local Instabilities with Tikhonov Regularization}
\label{sec:meth_regularize}
When random fiber networks deform under uniaxial tension, there is strong Poisson's effect that arises from the fibers oriented transverse to the loading direction buckling. As such, local instabilities (from beam buckling), and near rigid body motions (from Poisson's effect) results in an ill-conditioned tangential stiffness matrix. This leads to convergence difficulties in our numerical simulations. One of the common solutions to deal with numerical convergence in fiber networks is to use dynamic simulations coupled with damping to ensure that the kinetic energy is relatively small \citep{parvez2024methodological}, or use dynamic relaxation \citep{merson2024machine}.  In our case, to deal with the ill-conditioned tangential stiffness matrix, we add Tikhonov regularization \citep{tikhonov1977solutions} to the quasi-static finite element problem. While Tikhonov regularization is traditionally used to solve ill-posed problems, it has been applied to solve structural mechanics problems in topology optimization \citep{ramos2016filtering, zhang2017material} to heuristically deal with numerical instabilities that arises in the finite element problem. To apply Tikhonov regularization to our problem, we add a regularization term to the total energy in the system such that:

\begin{equation}
    \label{eqn:regularization}
    \Pi_{R} = \Pi_{int} + \int_{\ell} \left[ \frac{1}{2} \left( \Delta \bm{\tilde{u}} \cdot \bm{C}\Delta \bm{\tilde{u}} \right) \right] ds
\end{equation}

where $\bm{C} = \text{diag}[c_u, c_\theta]$ is the regularization parameter with $c_u$ regularizing the displacement degrees of freedom and $c_\theta$ regularizing the rotational degrees of freedom respectively, and $\Delta \bm{\tilde{u}} = [\bm{\tilde{u}-\tilde{u}_{prev}}]$, where $\tilde{\bm{u}} = [\bm{u},\theta]$ is the generalized displacements and $\tilde{\bm{u}}_{prev}$ is the generalized displacements from the previously converged solution. As such, the stationary conditions now become:

\begin{equation}
    \begin{aligned}
             & \Pi_{R}' \delta \bm{\tilde{u}} =  \int_{\ell} \left[ \delta\bm{\tilde{\varepsilon}} \cdot \bm{\tilde{\sigma}} + \delta \tilde{\bm{u}} \cdot \bm{C} \Delta\tilde{\bm{u}} \right] \; ds = 0 \\
             & \Pi_{R}' \delta \bm{\tilde\sigma} = \int_{\ell} \left[\delta \bm{\tilde{\sigma}} \cdot \big( \bm{\tilde{\varepsilon} - \mathbb{C}^{-1} \bm{\tilde{\sigma}}} \big) \right] \; ds = 0 \, .
        \end{aligned}
\end{equation}
To choose an appropriate value for the regularization parameter $\bm{C}$, we analyze the effects of the regularization term on the finite element problem in Section \ref{sec:meth_fea}.

\subsubsection{The Finite Element Formulation}
\label{sec:meth_fea}
To solve the finite element problem, we discretize the problem by introducing the nodal degrees of freedoms of the generalized displacement and  generalized stresses as $\bm{d}_u$ and $\bm{d}_\sigma$ respectively. As such, the displacements and stress fields can be approximated by:

\begin{equation}
\label{eqn:interp_func}
    \begin{aligned}
    \bm{\tilde{u}}_h &= \bm{N_ud}_u & \tilde{\bm{\varepsilon}}_h &= \bm{Bd}_u & \tilde{\bm{\sigma}}_h &= \bm{N_{\sigma}d_{\sigma}} \\ 
    \delta\bm{\tilde{u}}_h &= \bm{N_u \delta d}_u & \delta\tilde{\bm{\varepsilon}}_h &= \bm{Q\delta d}_u & \delta \tilde{\bm{\sigma}}_h &= \bm{N_{\sigma}\delta d_{\sigma}}
    \end{aligned}
\end{equation}

where the subscript $h$ denotes the discretized approximation and $\bm{N}_u$ and $\bm{N}_{\sigma}$ denote the appropriate interpolation functions for generalized displacements and stresses respectively. The previous displacement $\tilde{\bm{u}}_{prev}$ is interpolated with the same interpolation functions as the current displacement $\tilde{\bm{u}}$. Note that the discretized differential matrices $\bm{B}$ and $\bm{Q}$ are nonlinear with respect to $\bm{d}_u$ due to rotations present in the strain measures. We obtain the discretized stationary conditions by combining Eqn. \ref{eqn:stationary} and Eqn. \ref{eqn:interp_func}:

\begin{equation}
\begin{aligned}
    \{ \Pi_{R}' \delta \tilde{\bm{u}}\} _h & = \int_{\ell}\delta \bm{d}_u \bigl[ \underbrace{\bm{Q}^T \bm{N}_{\sigma}}_{\bm{\mathcal{Q^T}}}\bm{d}_\sigma + \underbrace{\bm{N}_{u}\bm{C}\bm{N}_{u}}_{\bm{\mathcal{M}}}\bm{d}_u - \bm{N}_{u}\bm{C}\bm{N}_{u}\bm{d}_{u_{prev}} \bigr] \; ds = 0\\
    \{ \Pi_{R}' \delta \tilde{\bm{\sigma}} \}_h & = \int_{\ell} \delta\bm{d}_{\sigma} \bigl[ \underbrace{\bm{N}_{\sigma}\bm{B}}_{\bm{\mathcal{B}}}\bm{d}_u - \underbrace{\bm{N}_{\sigma}^T\mathbb{C}\bm{N}^T_{\sigma}}_{\bm{\mathcal{C}}}\bm{d}_\sigma \bigr] \; ds = 0 \, .
\end{aligned}
\end{equation}
Since the stationary conditions shown above must hold for all admissible test functions $\delta\bm{d}_u$ and $\delta\bm{d}_\sigma$, the discretized stationary conditions can be written in matrix form as:

\begin{equation}
\label{eqn:discretized_stationary}
    \begin{bmatrix}
        - \bm{\mathcal{C}}  & \bm{\mathcal{B}} \\
        \bm{\mathcal{Q}}^T & \bm{\mathcal{M}}
    \end{bmatrix}
    \begin{bmatrix}
        \bm{d}_{\sigma} \\ 
        \bm{d}_u
    \end{bmatrix}
     = 
     \begin{bmatrix}
        \bm{0} \\
        \bm{\mathcal{\bm{M}}}\bm{d}_{u_{prev}}
     \end{bmatrix} \, .
\end{equation}
To solve this nonlinear system of equations, we use the Newton-Raphson method. Following standard procedures~\citep{belytschko2014nonlinear, magisano2017improve}, we linearize the stationary conditions using the first-order Taylor series expansion. As a result, the linearized stationary conditions can be written as:

\begin{equation}
\begin{aligned}
    \{ \Pi_{R}'\delta\tilde{\bm{u}}\}_h & \approx \bm{R}_u + \int_{\ell} \bigl[ \delta \dot{\tilde{\bm{\varepsilon}}} \cdot \tilde{\bm{\sigma}}+ \delta\tilde{\bm{\varepsilon}} \cdot \dot{\tilde{\bm{\sigma}}} + \delta \tilde{\bm{u}} \cdot C \dot{\tilde{\bm{u}}} \bigr] \; ds = 0 \\
    \{ \Pi_{R}'\delta\tilde{\bm{\sigma}} \}_h & \approx \bm{R}_{\sigma} + \int_{\ell} \bigl[ \delta \tilde{\bm{\sigma}} \cdot ( \dot{\tilde{\bm{\varepsilon}}} - \mathbb{C}^{-1} \cdot \dot{\tilde{\bm{\sigma}}} )\bigr] \; ds = 0
\end{aligned}
\end{equation}
where $\bm{R}_{u_h} = \bm{\mathcal{Q}}^T\bm{d}_{\sigma} + \bm{\mathcal{M}}(\bm{d}_u-\bm{d}_{u_{prev}})$ and $\bm{R}_{\sigma_h} = -\bm{\mathcal{C}}\bm{d}_\sigma + \bm{\mathcal{B}}\bm{d}_u$ are the residuals from the previous Newton iterations (see Eqn. \ref{eqn:discretized_stationary}) and $\dot{(.)}$ denotes the incremental change between Newton iterations. Again, we note that discretizing $\delta \dot{\tilde{\bm{\varepsilon}}}$ is nontrivial due to rotations present in the strain measures. Here we let $\delta \dot{\tilde{\bm{\varepsilon}}}_h = \dot{\bm{d}}_u\mathcal{\bm{\psi}} \delta \bm{d}_u$ with the explicit form for $\mathcal{\bm{\psi}}$ can be found in \citep{ritto2002differentiation}. As a result, the discretized linearized stationary condition equations are:

\begin{equation}
\begin{aligned}
    \{ \Pi_{R}'\delta\tilde{\bm{u}}\}_h & \approx \bm{R}_u + \int_{\ell} \bigl[ \delta \bm{d}_u^T \bm{\psi}^T \dot{\bm{d}}_u \bm{N}_{\sigma}\bm{d}_{\sigma} + \delta \bm{d}_u^T \bm{\mathcal{Q}}^T \bm{N}_{\sigma}\dot{\bm{d}}_\sigma + \delta\bm{d}_u^T\bm{N}_u^T C\bm{N}_u\bm{d}_u\bigr] \; ds = 0 \\
    \{ \Pi_{R}'\delta\tilde{\bm{\sigma}} \}_h & \approx \bm{R}_{\sigma} + \int_{\ell} \bigl[ 
\delta\bm{d}_\sigma^T \bm{N}_\sigma^T (\bm{\mathcal{Q}}\dot{\bm{d}}_u - \mathbb{C}^{-1}\bm{N}_\sigma\dot{\bm{d}}_\sigma)  \bigr] \; ds = 0 \, .
\end{aligned}
\end{equation}
Similarly, for all linearized stationary conditions to hold for all admissible test functions, the linearized stationary conditions can be written as:
\begin{equation}
\label{eqn:Newton}
\begin{bmatrix}
- \bm{\mathcal{C}} & \bm{\mathcal{Q}} \\
\bm{\mathcal{Q}}^T & \bm{\mathcal{G}} + \bm{\mathcal{M}}
\end{bmatrix}
\begin{bmatrix}
\dot{\bm{d}}_{\sigma} \\ \dot{\bm{d}}_u 
\end{bmatrix}
=
- \begin{bmatrix}
\bm{R}_{\sigma_h} \\ \bm{R}_{u_h}
\end{bmatrix}
\end{equation}
where $\mathcal{G} = \psi\bm{N}_\sigma \bm{d}_{\sigma}$.
The above matrix equation is iteratively solved until convergence.

To understand and analyze how the hybrid beam formulation and Tikhonov regularization affect the convergence of the Newton iterations, we perform static condensation of $\dot{\bm{d}}_\sigma$ on Eqn. \ref{eqn:Newton} to gain an equivalent tangential stiffness matrix $\bm{K}$:

\begin{equation}
\label{eqn:static_condense}
\Big(\underbrace{ \bm{\mathcal{Q}}^T\bm{\mathcal{C}}^{-1}\bm{Q} + \bm{\mathcal{G}} + \bm{\mathcal{M}}}_{\bm{K}} \Big) \dot{\bm{d}}_{u} = -\bm{R}_{u_h}-\bm{\mathcal{Q}}^T \bm{\mathcal{C}}^{-1}\bm{R}_{\sigma_h} \, .
\end{equation}

Here, the equivalent stiffness matrix $\bm{K}$ is composed of three terms. The first two terms (i.e., $\bm{\mathcal{Q}}^T\bm{\mathcal{C}}^{-1}\bm{\mathcal{Q}}$ and $\bm{\mathcal{G}}$) are the tangential material and geometric stiffness matrix respectively. While the equations may look similar to traditional displacement based finite elements, we emphasize that the stresses are computed and interpolated independently from the displacements. Specifically, the geometric stiffness matrix $\mathcal{\bm{G}} = \mathcal{\bm{G}}[\bm{d}_{\sigma}, \bm{d}_u]$ in the case of hybrid beam elements, but $\bm{\mathcal{G}} = \bm{\mathcal{G}[\tilde{\bm{\varepsilon}}[\bm{d}_u], \bm{d}_u}]$ in the case of displacement-based beam elements since $\tilde{\bm{\sigma}}_h = \mathbb{C} \tilde{\bm{\varepsilon}}_h = \mathbb{C} \bm{B} \bm{d}_u$ for the displacement-based formulations. In the case of large deformation and small strains, for displacement based finite elements, $\tilde{\bm{\sigma}}_h$ can be a poor estimate of the stresses on the equilibrium curve since $\tilde{\bm{\sigma}}_h$ is a nonlinear function with respect to $\bm{d}_u$. In the case of hybrid elements, since $\tilde{\bm{\sigma}}_h$ is interpolated and solved independently,  $\tilde{\bm{\sigma}}_h$ is not constrained to satisfy the constitutive relations during Newton iterations. Thus, the error between the $\tilde{\bm{\sigma}}_h$ and the stress at the equilibrium points is reduced significantly. More formal and rigorous analysis on hybrid/mixed methods in the context of slender structures and extrapolation locking can be found in the literature \citep{magisano2017advantages}. 

\begin{algorithm}[h]
    \DontPrintSemicolon
    \label{alg:solver}
    \caption{Convergence Criteria for Random Fiber Networks}
    
    \textbf{Initialize:} $\varepsilon_{max}, \; c_0, \; tol, \; count_{max}$ \;
    
    \BlankLine
    converged $\leftarrow$ False \; 
    \While{converged == False}{
    Newton Solve \tcp{Get converged Newton Solution}
    Compute $\frac{R_{eq}}{F_{reac}}$ \tcp{Compute additional convergence criteria}
    count $\leftarrow 0$ \;
    \BlankLine
    \eIf(\tcp*[h]{Case where residual is over tolerance}){$\frac{R_{eq}}{F_{reac}} \geq \text{tol}$}{
    $\tilde{\bm{u}}_{prev} = \tilde{\bm{u}}$ \tcp{Update previous converged solution for regularization}
    
    \BlankLine
    \If(\tcp*[h]{For cases of slow equilibrium convergence}){count $> count_{max}$}
    {
    $c_0 \leftarrow \frac{c_0}{2}$\tcp{Reduce the regularization strength}
    count $\leftarrow 0$
    }
    }
    {converged $\rightarrow$ True \tcp{Case where tolerance is met; Solution converged}}
    }
    Increment to next solver step (i.e., increment applied force/displacement)
    
    \end{algorithm}

Finally, the last term in $\bm{K}$ (i.e., $\bm{\mathcal{M}}$) comes from Tikhonov regularization. Note that in our case the regularization term has the form of a mass matrix. Since mass matrices are positive definite \citep{hughes2003finite}, it ensures that $\bm{K}$ is well-conditioned assuming proper selection of $\bm{C}$. However, in addition to modifying the equivalent stiffness $\bm{K}$, a large $\bm{C}$ term also effects the residual calculation (see Eqn. \ref{eqn:discretized_stationary}). In our work, to make sure that $\bm{C} = \text{diag}[c_u, c_\theta]$ is small enough such that it doesn't affect the residual calculation but still sufficiently large enough to keep $\bm{K}$ positive definite, we heuristically let $c_u = (c_0/M_u) \text{tr}[\bm{K}-\bm{\mathcal{M}}]_u$ and $c_\theta = (c_0/M_\theta) \text{tr}[\bm{K}-\bm{\mathcal{M}}]_\theta$ where $c_0 = 10^{-7}$, $M$ denotes the size of the stiffness matrix, $\text{tr}[\dots]$ denotes the matrix trace operator and the subscripts $u$ and $\theta$ denotes the displacement portion of the matrix and rotation portion of the matrix respectively. To ensure that equilibrium is still preserved, we have a convergence criterion (in addition to the Newton solver convergence criterion) applied to the ratio between the equilibrium residual and reaction force $||\bm{R}_{eq}||/F_{reac} \leq 10^{-4}$, where $\bm{R}_{eq} = [\bm{\mathcal{Q}}^T\bm{d}_\sigma$, $\bm{R}_\sigma$], $||\dots||$ denotes the $L_2$ norm, and $F_{reac}$ is the total reaction force from the applied displacement. The additional convergence criterion also facilitates the tuning of the regularization strength, with details of our convergence criterion and regularization tuning shown in Algorithm \ref{alg:solver}. For readers interested in our implementation, we point to our Github page \url{https://github.com/pprachas/struc_func_fiber_network} for specific solver details (e.g., adaptive Newton step-size, Newton convergence criteria, etc.). \changes{Additional details on our implementation of Tikhonov regularization and sensitivity analysis of the finite element solution to the additional convergence criterion are also provided in Appendix \ref{appendix:regularization}.}
In this work, we made sure that our solution is agnostic to mesh density before generating our results. Mesh refinement studies  on representative domains used in this work can be found in Appendix \ref{appendix:mesh_refinement}.

\subsection{Analytical Reduced Order Model Connecting Geometry to Mechanics}
\label{sec:meth_rom}

In addition to the numerical simulations described in Section \ref{sec:meth_model}, we derive an analytical reduced order model of these fibers. The motivation for creating this analytical model is twofold. First, as demonstrated in Section \ref{sec:meth_model}, numerically simulating these systems is non-trivial. Therefore, agreement between an analytical model and simulation results will provide stronger support for our mechanistic conclusions. Second, the computational model provides limited insight into the relationship between the heterogeneous geometry of random fiber networks and their mechanical behavior. Thus, complementing simulations with a simple analytical model would add mechanistic insight. In this Section, we derived a simple analytical model connecting the geometry of random fiber networks to random fiber network mechanics. We begin in Section \ref{sec:meth_rom_single} with the derivation of our analytical reduced order model for an arbitrary fiber chain. Then, with the analytical reduced order model for a single chain in hand, we extend the model to capture the behavior of fiber networks as described in Section \ref{sec:meth_rom_network}.

\subsubsection{Analytical Reduced Order Model for Single Fiber Chains}
\label{sec:meth_rom_single}

\begin{figure}[t]
\includegraphics[width= \textwidth]{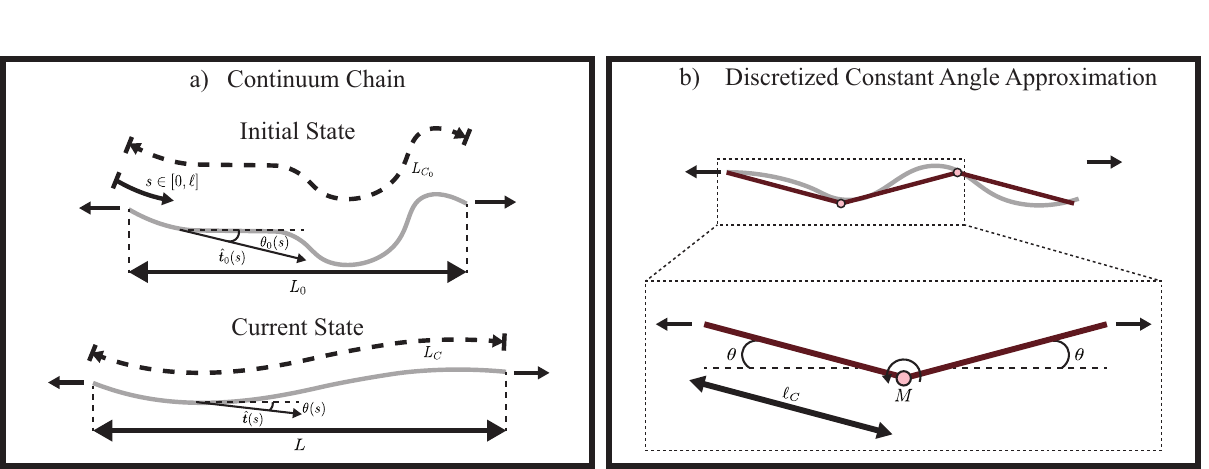}
\caption{Schematic of our analytical model describing the mechanics of single fiber chains. a) Illustration of the kinematics of continuous chains in the initial and current configuration; b) Outline of the discretized chain approximation used in our rotational spring model. Note that $\theta$ is constant for all links in the discretized approximation.}
\label{fig:meth_rom}
\end{figure}

We begin by describing the kinematics of a flexible and extensible chain with arbitrary shape under uniaxial tension (see Fig. \ref{fig:meth_rom}a). Similar to the geometrically exact beams introduced in Section \ref{sec:meth_2D}, we parametrize the chain by $s \in [0,\ell]$ in the initial undeformed configuration where $\ell$ is the total beam length. The position of the chain in the initial reference configuration is denoted by the vector $\bm{r}_{C_0}(s)$, and the position of the chain in the current configuration is denoted by $\bm{r}_C(s)$. The tangent vector in the initial configuration is then $\bm{r}_{C_0,s}(s) = \bm{t}_{C_0}(s)$, and the tangent vector in the current configuration is then $\bm{r}_{C,s}(s) = \bm{t}_C(s)$ where $(.)_{,s} = \frac{d (.)}{ds}$. As such, the total contour length of the chain can be written as: 

\begin{equation}
\begin{aligned}
L_{C_0} &= \int_\ell ||\bm{t}_{C_0}(s)|| \; ds  \\ 
L_C &=  \int_\ell ||\bm{t}_C(s)|| \; ds 
\end{aligned}
\end{equation}

where $L_{C_0}$ and $L_C$ are the total contour length in the initial and current configurations respectively. Finally, the end-to-end length of the chain can be computed as:

\begin{equation}
\begin{aligned}
    L_0 = &\int_\ell ||\bm{t}_{C_0}(s)|| \, \cos \theta_{C_0}(s) \; ds \\
    L = &\int_\ell ||\bm{t}_C(s)|| \, \cos \theta_C(s) \; ds
\end{aligned}
\end{equation}

where $\theta_{C_0}(s)$ and $\theta_C(s)$ are the angles between the tangent vector with the loading direction in the initial and current states respectively.

To describe the mechanics of the chain, we first define the chain cross-sectional area as $A$, the second moment of area as $I$, and Young's modulus as $E$. We then construct a simple analytical reduced order model through the linear combination of a stretching spring model and a rotational spring model. Starting from the stretching spring model, the force required to deform a linear spring can be written as:

\begin{equation}
    F_s=k_s \Delta x
\end{equation}

where $k_s$ is the linear spring stiffness and $\Delta x$ is the change in length of the spring. From structural mechanics, the equivalent stiffness of the chain is $k_s = \frac{EA}{L_{C_0}}$ and the change in length can be written in terms of the contour lengths such that $\Delta x = L_C-L_{C_0}$. As such, the reaction force from the stretching component can be written as:

\begin{equation}
F_s = \frac{EA}{L_{C_0}} \Biggl( L_C-L_{C_0} \Biggr) \, .
\end{equation}

Following the linear stretching term, we also use a simple linear rotational spring that can be expressed as:

\begin{equation}
M = k_\theta \Delta \theta
\end{equation}

where $k_\theta$ is the rotational spring stiffness and $\Delta \theta = \theta_C - \theta_{C_0}$ is the change in angle. Again, from structural mechanics, the equivalent torsion spring stiffness is $k_\theta = \frac{\alpha EI}{\ell_C}$ where $\ell_C$ is some characteristic length and $\alpha$ depends on the stiffness modification factor from boundary conditions \citep{alderliesten2018introduction}. However, computing $\Delta \theta$ is less straight-forward. To compute $\Delta \theta$, we first start by looking at the the initial tortuosity $\tau_0$ of the structure which we define as:

\begin{equation}
\label{eqn:tortuosity}
\tau_0 = \frac{L_{C_0}}{L_0} = \frac{\displaystyle\int_\ell ||\bm{t}_{C_0}(s)|| \; ds}{\displaystyle\int_\ell ||\bm{t}_{C_0}(s)|| \, \cos \theta_{C_0}(s) \; ds} \, .
\end{equation}

\changes{Here we make the assumption that $\theta_{C_0}(s) = \theta_{C_0}$ is a constant, meaning that the term $\cos \theta_{C_0}$ can be factored out of the integral. While this might seem extremely constraining for an arbitrary continuum chain (as illustrated in Fig. \ref{fig:meth_rom}a) this condition is significantly less restricting in practice, i.e., in the case of discrete chains such as the ones that make up a fiber network (see Fig. \ref{fig:meth_rom}b).} This crucial bending spring model assumption implies that the continuous chain can be approximately discretized into inextensible rods of length $\ell_C$ with constant $\theta_{C_0}$ (see Fig. \ref{fig:meth_rom}b). Furthermore, with $\theta_{C_0}$ defined as a constant, $\theta_{C_0}$ can be calculated as:

\begin{equation}
\label{eqn:approx_angle}
\theta_{C_0} = \arccos \left( \frac{L_0}{L_{C_0}} \right) = \arccos \left( \tau_0^{-1} \right) \;.
\end{equation}

Following the same logic, computation of the angle in the current configuration can be written as $\theta_C = \arccos \left( \frac{L}{L_C} \right)$. As such, the force required to induce this reaction moment in the current chain configuration can be computed from the discretized approximation of the chain as (see Fig. \ref{fig:meth_rom}b):

\begin{equation}
F_r = \frac{M}{\ell_C} \sin \theta_{C} \, .
\end{equation}

The two stretching and bending terms are combined to get the total reaction force such that:

\begin{equation}
F = F_s + F_r \, .
\end{equation}

To facilitate nondimensionalization, we introduce a dimensionless parameter $\tilde{\kappa}$, also referred to as the dimensionless bending length, that can be written as:

\begin{equation}
\label{eqn:kappa_tilde}
\tilde{\kappa} = \frac{EI}{EA\ell_C^2} \, .
\end{equation}

Analogous to the persistence length in worm-like chain models, $\tilde{\kappa}$ is shown to Sbe critical in determining the mechanics of random fiber networks \citep{head2003distinct, picu2011mechanics, wilhelm2003elasticity}. Note that $\tilde{\kappa}$ is a pure geometric parameter and can be written as $\left(\frac{R_g}{\ell_C}\right)^2$ where $R_g = \sqrt{\frac{I}{A}}$ is the radius of gyration. As such, $\tilde{\kappa}$ can also be interpreted as a measure of the slenderness of the chain where a smaller $\tilde{\kappa}$ value leads to more slender fibers. To conform to biological and natural systems, in this work we will investigate systems with $\tilde{\kappa} \in \{ 10^{-6},10^{-5},10^{-4},10^{-3} \}$ \citep{sharma2016strain}.

Finally, after nondimensionalization, the analytical reduced order model is expressed as:

\begin{equation}
\label{eqn:meth_rom}
\underbrace{\frac{F}{EA}}_{\text{Reaction Force}} \quad = \quad \frac{\sigma_{yy}}{E} \quad = \quad \biggl( \underbrace{\frac{L_C}{L_{C_0}} - 1 }_{\text{Stretching Spring}} \biggr) \quad + \quad \biggl( \underbrace{\alpha \tilde{\kappa}\Delta\theta \vphantom{\frac{L_C}{L_{C_0}}}\sin\theta_{C}}_{\text{Rotational Spring}} \biggr) \, .
\end{equation}

\subsubsection{Analytical Reduced Order Model for Random Fiber Networks}
\label{sec:meth_rom_network}

\begin{figure}[t]
    \centering
    \includegraphics[width= 0.95\textwidth]{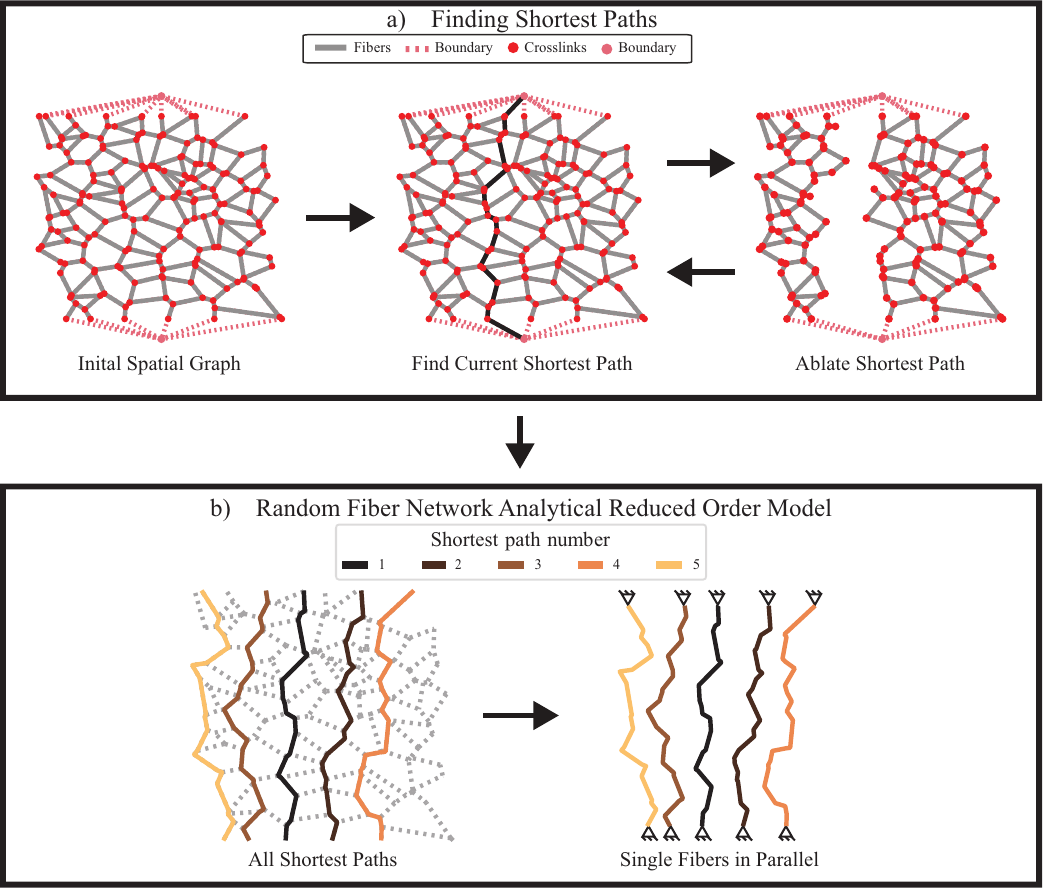}
\caption{Flowchart to apply the analytical reduced order model in Section \ref{sec:meth_rom_single} to random fiber networks. a) Illustrates the pipeline to extract the predicted force chains from our random fiber networks. We first convert our random fiber network into a spatial graph and connect the constrained boundaries to a boundary node. We then find the shortest path between the boundaries and ablate the shortest path to form a new spatial graph. This process is repeated until no path remain between the constrained boundaries. The collection of shortest paths extracted are our predicted force chains location. b) Outline of applying the analytical reduced order model our predicted force chains. We apply our derived analytical reduced order model to each path separately and combine the paths into one system by interpreting the set of paths as springs in parallel.}
\label{fig:meth_fibers}

\end{figure}

Given our analytical reduced order model for single fiber chains in Section \ref{sec:meth_rom_single}, our next goal is to extend the analytical reduced order model to predict the mechanics of random fiber networks. It is not necessarily obvious how to directly apply the reduced order model to random fiber networks. However, the emergence of force chains, or load paths, in random fiber networks under loading suggests that only a subset of fibers in the fiber network will contribute to mechanical behavior \citep{parvez2023stiffening, sarkar2022evolution}. In the broader literature, force chains are also studied in granular systems \citep{guerra2022elastogranular, karapiperis2021nonlocality, peters2005characterization, tordesillas2010force} where there is interest in understanding the emerging topological structure of loading paths that form during granular jamming. In both the case of random fiber networks and granular systems, identification of force chains is typically an ad-hoc process that requires information about the force distribution in the network. \changes{In this work, we present a pipeline to 1) predict the location of force chains from the geometric structure of fiber networks alone, and 2) combine force chains with our analytical reduced order model to approximate the mechanics of random fiber \textit{networks}. In combination, these two goals aim to map the structure of random fiber networks to their mechanical function.}

Our process for predicting force chains from random fiber network geometry is illustrated in Fig. \ref{fig:meth_fibers}. 
First, we convert the fiber network into a graph $G = (V,E,\ell_N)$ where the set of vertices (or nodes) $V=\{v_1, \dots, v_k \}$ are the fiber-fiber crosslinks, the set of edges $E \subseteq V \times V$ are the fibers themselves, and $\ell_N:E \rightarrow \mathbb{R}$ is the weight of each edge which is prescribed to be the Euclidean distance between the vertices of each edge (i.e., length of the fiber). 
Here we propose the \textit{distance-weighted shortest path between boundaries} as the geometric feature that governs the mechanical response of random fiber networks. For context, a path in graph theory is a set of adjacent edges that joins a sequence of vertices~\citep{west2001introduction}. A path with no repeating vertices, and consequently no repeating edges, is called a simple path. For brevity, any path mentioned in this work is a simple path unless specified otherwise. 

For an arbitrary path $\eta \subset E$ with sequence of edges $\eta_{i,j}$, the total distance-weighted path is defined as:

\begin{equation}
D(\eta) = \sum_{i,j} \ell_N(\eta_{i,j}) \, .
\end{equation}

To compute the set of shortest paths $\Phi = \{ \phi^1,\phi^2, \dots, \phi^k \}$ where $\phi^k$ is the sequence of edges that form the shortest path, we first construct $2$ boundary nodes that correspond to the constrained boundary conditions (see Fig. \ref{fig:meth_fibers}a). We then connect the fiber nodes that are fixed to the boundary nodes through a boundary edge with weight zero. We note that the boundary nodes and edges in Fig. \ref{fig:meth_fibers}a are for visualization purposes only; thus, illustrated boundary edge distances do not actually represent distances in Euclidean space. The shortest path is then computed between the top and bottom boundary using Dijkstra's shortest path algorithm \citep{dijkstra2022note} through the NetworkX library \citep{hagberg2008exploring}. After computing the shortest path, the path is ablated from the fiber network and the process of finding the shortest paths is repeated until no path between the constrained boundaries remains. Note that in the case where there is only one possible path left between the constrained boundaries, that path automatically becomes the shortest path. The collection of shortest paths $\Phi$ predicted by this algorithm then becomes our predicted location of force chains. By construction, $D(\phi^1) \leq D(\phi^2) \dots \leq D(\phi^k)$ (i.e., the shortest path number is arranged from least total distance to largest total distance). The mechanics of each individual path can then be inferred through our derived analytical reduced order model from Section \ref{sec:meth_single}, and merged into one system by analyzing the set of paths as spring in parallel.

\subsection{Geometric Parameter Space and Boundary Conditions}
\label{sec:meth_geo}
In this Section, we introduce the geometric domains that we perform our analysis on. The pipeline to generate our single fiber chains as well as applied boundary conditions is outlined in Section \ref{sec:meth_single}. In Section \ref{sec:meth_network}, we detail the pipeline to generate the random fiber networks and specify the boundary conditions.

\begin{figure}[ht]
    \centering
    \includegraphics[width= \textwidth]{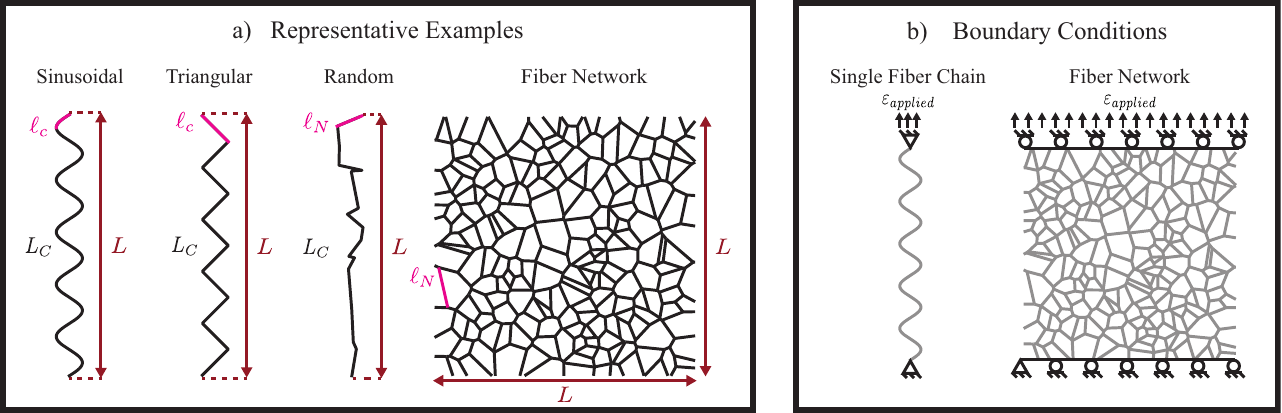}
    \caption{Visualization of representative examples of structures that we are investigating in this work as well as the boundary condition of these structures. a) Illustration of representative examples of structures used in this study and the main geometric parameters important to our analysis. b) Boundary conditions for single chains and fiber networks. The single fiber chains are constrained with pinned-pinned boundary conditions while fiber networks are constrained with roller-roller supports. In the case of fiber networks, one node is constrained in the vertical direction to avoid rigid body motion. Both structures are subjected to applied displacement as external loading. }
    \label{fig:meth_examples}
\end{figure}

\subsubsection{Single Fiber Chain Generation}
\label{sec:meth_single}
To make sure that our approach can be applied to a diverse set of single fiber chains, we perform computational experiments on $3$ different chain types: sinusoidal chains, triangular chains, and random chains. Examples of these chains are shown in Fig. \ref{fig:meth_examples}a. For periodic chains (i.e., sinusoidal and triangular chains), the geometry of the chains can be represented by:

\begin{equation}
    \label{eqn:generate_periodic}
    x = a \; f \left[ \frac{2\pi y}{\lambda} \right] \quad 
\end{equation}

where $y \in [0,L_0]$ is the domain, $a \in \{ L_0/40, L_0/20, L_0/10 \}$ is the amplitude, $f[\dots]$ is the periodic function (i.e., sinusoidal or triangular wave function) and $\lambda \in \{ L_0/40, L_0/20, L_0/10, L_0/5, L_0 \}$ is the wavelength. For random chains, we first obtain $n_c = \{ 10, 20, 30, 40, 50, 60, 70, 80 \}$ crosslinks by sampling ${y_1, \dots, y_{n_c-1}} \sim \mathcal{U}(0,L_0)$ and ${x_1, \dots, x_{n_c-1}} \sim \mathcal{U}(-a/2, a/2)$ where $\mathcal{U}$ is the uniform distribution. From the sampled crosslinks, fiber chains are created by connecting the closest adjacent points while maintaining monotonic relationship in $y$ to prevent intersecting fibers. This set of parameters for each type of chain produces fibers with geometric dimensions that varies from random fiber networks in the mesoscale range (i.e., where boundary effects and may play a role) to random fiber networks that can be represented in the macroscale range (i.e., length-scale separation between fiber size and system size) \citep{merson2020size, shahsavari2012model, shahsavari2013size}. We note that the crosslinks are considered to ``weld'' the two adjacent fibers (i.e., displacements and rotations of the two fibers are constrained to be the same at the crosslinks). Additionally, for consistent boundaries, we let $(x_0,y_0) = (0,0)$ and $(x_n,y_n) = (0,L_0)$. Without loss of generality, in this work, we define $L_0=10000$. The characteristic length for a chain with $N$ fibers is $\ell_c = (1/N) \sum_{N} \ell_N$ with $\ell_N$ being the length of each fiber in the chain. Note that in the case of periodic chains, we let $\ell_C = \lambda/2$.  Finally, we use pinned-pinned supports with incrementally applied global strain $\varepsilon_{applied}$ as boundary conditions for single fiber chains (see Fig \ref{fig:meth_examples}). As such, $\alpha = 3$ in our analytical reduced order model (see Eqn. \ref{eqn:meth_rom}). 

\subsubsection{Fiber Network Generation}
\label{sec:meth_network}
\newchanges{For random fiber networks to have a strain-stiffening effect from fiber kinematics akin to those typically observed in biological tissue, the small-strain behavior must be bending dominated \citep{picu2021constitutive,sharma2016strain}. The necessary conditions separating bending dominating and stretching dominated microstructures is based on Maxwell's isostatic stability criterion \citep{maxwell1864calculation} for pin jointed structures. If the structure's domain geometry with pin joints have rigid body modes, then the same structure with welded joints will be bending dominated under small deformations \citep{fleck2010micro}. In the case of large structures like random fiber networks, the network connectivity (i.e., the average number of fibers connected to a crosslink) $\langle z \rangle \leq 4$ is a necessary \emph{but not sufficient} condition for 2D networks to be bending dominated \citep{licup2015stress, sharma2016strain}, as structures with large $\tilde{\kappa}$ and high density are stretching dominated even for small strains \citep{picu2011mechanics}.} 

To generate in our random fiber networks that are under the isostatic limit geometrically, we use 2D Voronoi diagrams which have $\langle z \rangle \approx 3$. Specifically, in this work, we generate random fiber networks of window size $L_0 \times L_0$ by first sampling $n \in \{100,200,300,400,500 \}$ initial seeds with $(x,y)_n \sim \mathcal{U}(0,L_0)$ and generate Voronoi diagrams from these initial seeds. The edges of the Voronoi diagram are the fibers and the vertices are the crosslinks. We note that similar to the crosslinks in single fiber chains, the crosslinks are ``welded''. Fibers that cross the $L_0 \times L_0$ window are cropped since they do not influence the mechanical response. An example of a Voronoi diagram with $n=200$ can be found in Fig. \ref{fig:meth_examples}a. Similar to single chains, we also set $L_0=10000$. Similarly, the characteristic length for a random fiber network with $N$ fibers is $\ell_c = (1/N) \sum_{N} \ell_N$, with $\ell_N$ being the length of each fiber in the fiber network. The boundary conditions for fiber networks are roller-roller constraints with one degree of freedom fixed in the horizontal direction to prevent rigid body motion (see Fig. \ref{fig:meth_examples}b). Again, $\alpha = 3$ is used for the analytical reduced order model (see Eqn.\ref{eqn:meth_rom}).  Similar to single fiber chains, we incrementally apply global strain $\varepsilon_{applied}$ as external loading. 

\section{Results and Discussion}
\label{sec:res}
In this work, our goal is to link the geometry structure and kinematics of random fiber networks to their mechanical function. To this end, we first analyze a simpler system in Section \ref{sec:res_mech_single}: the single fiber chain. Then, after building structure-function relationships for single fiber chains, we extend the analysis to random fiber networks in Section \ref{sec:res_mech_fibers}.

\subsection{Mechanics of Single Fiber Chains}
\label{sec:res_mech_single}

\begin{figure}[ht]
    \centering
    \includegraphics[width= \textwidth]{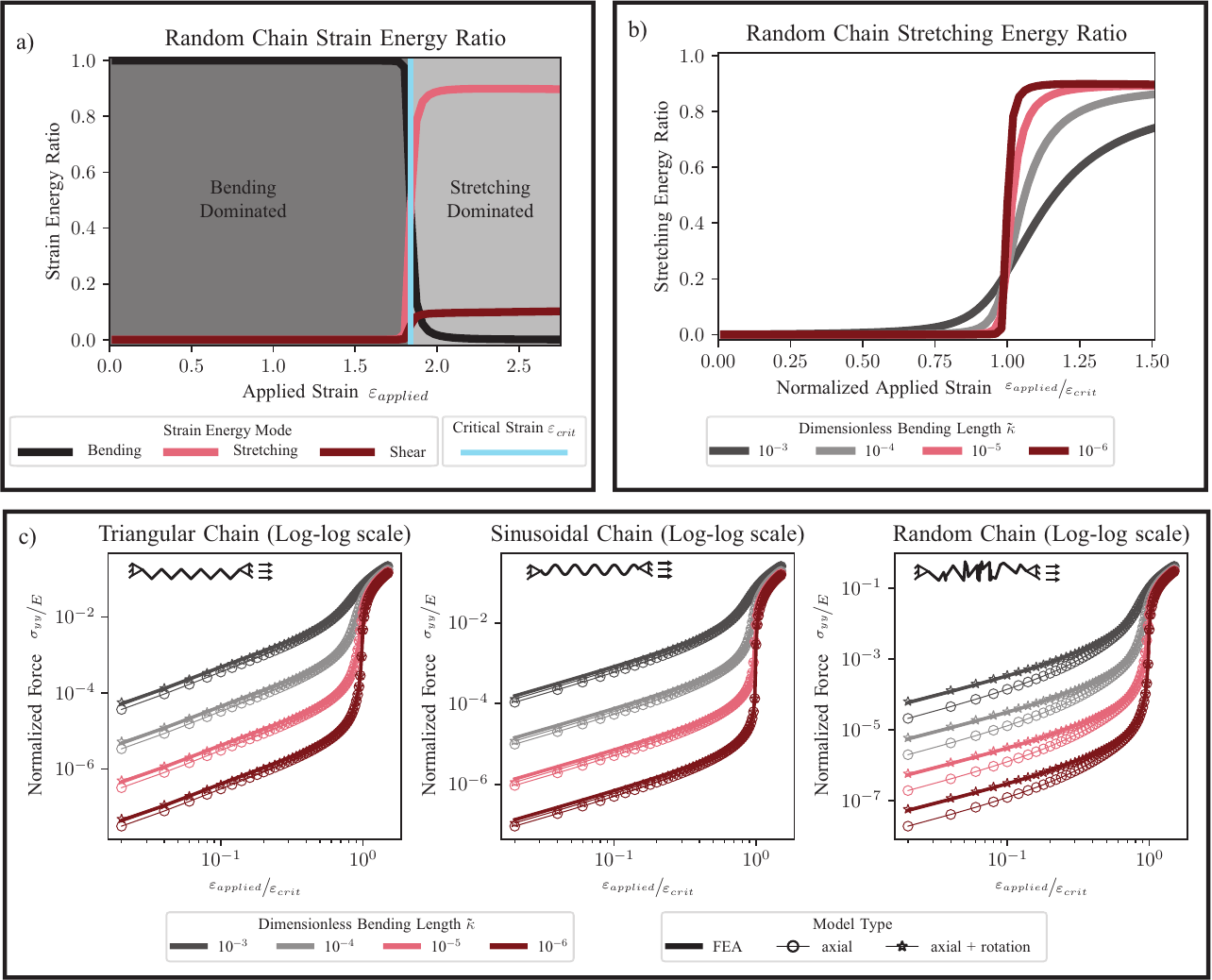}
    \caption{Representative examples to demonstrate the mechanics of single fiber chains. a) Strain energy distribution of single fiber chains under uniaxial tension. The representative example shown is from a random chain with $\tilde{\kappa}=10^{-6}$. Initially, the fiber chain is unraveling and is under a stretching dominated regime. Once the fiber chain is sufficiently straightened, the fiber transitions from a bending dominated deformation mode to a stretching dominated mode at a critical strain value $\varepsilon_{crit}$. The shear deformation mode remains small during the whole duration of loading. b) Comparisons of stretching energy distribution for different $\tilde{\kappa}$. The transition between the stretching and bending deformation mode becomes sharper as $\tilde{\kappa}$ decreases (i.e., for more slender fibers). c) Comparisons between the performance of our analytical reduced order model and FEA results on predicting the dimensionless reaction force $\sigma_{yy}/E$ for all $3$ types of single fiber chains. Visualization of the undeformed chain is shown on the top left of the figure. Note that the plots are in log-log scale.}
    \label{fig:res_single_mech}
\end{figure}

In this Section, we examine the mechanics of single fiber chains under uniaxial loading. Since our single fiber chains are slender (i.e., $\tilde{\kappa} \in [10^{-6},10^{-3}]$; see Eqn. \ref{eqn:kappa_tilde} for geometric interpretation of $\tilde{\kappa}$), the interaction between dominant deformation modes determines the mechanics of these chains  \citep{holmes2019elasticity}. Specifically, the deformation modes present in our computational model of single fiber chains are stretching deformation, shear deformation and bending deformation (see Eqn. \ref{eqn:beam_int}). As such, the strain energy of these single fiber chains can be decomposed into stretching energy $\Pi_{stretch}$, shearing energy $\Pi_{shear}$ and bending energy $\Pi_{bend}$. We analyze the interaction of the deformation modes under loading by looking at the change of energy ratio under loading, with the energy ratio being defined as:

\begin{equation}
 \label{eqn:enery_ratio}
 \frac{\Pi_{[\dots]}}{\Pi_{stretch}+\Pi_{shear}+\Pi_{bend}}
\end{equation}

where $\Pi_{[\dots]}$ is the energy mode we are interested in (i.e., $\Pi_{stretch}$, $\Pi_{shear}$, or $\Pi_{bend}$).
The energy ratio under loading of a representative example of a random single fiber chain (parameters $n_c = 30$, $a=L_0/10$, $\tilde{\kappa} = 10^{-6}$) is illustrated in Fig. \ref{fig:res_single_mech}a. Note that since the chain is slender, the contribution from the shear deformation mode is significantly lower than the contribution from the bending and stretching modes. Initially, as strain $\varepsilon_{applied}$ is applied, the chain is in the bending energy dominated regime since it is energetically more favorable for slender structures to bend. Physically, the fiber chain is unraveling (i.e., straightening out) in the bending dominated regime. As the fiber chain starts to straighten, the fiber transitions from a bending energy dominated regime to a stretching energy dominated regime at a critical transition strain $\varepsilon_{crit}$. In this work, $\varepsilon_{crit}$ is computed and defined by identifying $\varepsilon_{applied}$ such that $\Pi_{bend} = \Pi_{stretch}$ via the bisection method. \newchanges{Note that this definition of $\varepsilon_{crit}$ is different from the ``onset of stiffening'' described in multiple places in the literature \citep{sarkar2022evolution,vzagar2015two}, where the ``onset of stiffening'' typically refers to the transition from linear elastic behavior  to nonlinear elastic behavior. In our case, $\varepsilon_{crit}$ is consistent with transition point to from exponential stiffening to power law stiffening (i.e., the transition from ``regime II'' to ``regime III'') reported in literature \citep{parvez2023stiffening,picu2021constitutive,vzagar2015two}. } From Fig. \ref{fig:res_single_mech}b, the rate of this transition depends on $\tilde{\kappa}$ where a chain with larger $\tilde{\kappa}$ (i.e., a thicker fiber) will have a more gradual transition between stretching and bending energy compared to a chain with smaller $\tilde{\kappa}$ (i.e., a thinner fiber). This transition from a bending energy dominated mode to a stretching energy dominated mode results in a strain-stiffening effect in the normalized force-displacement curve as visualized from the representative examples (sinusoidal and triangular chain with parameters $\lambda = L_0/5$, $a = L_0/20$ ; random chain with parameters $n_c = 30$, $a=L_0/10$) in Fig. \ref{fig:res_single_mech}c. Note that in this work, we report the reaction force as a dimensionless value $\sigma_{yy}/E$ for both the finite element model and the analytical model (see Eqn. \ref{eqn:meth_rom} for dimensionless form of our analytical reduced order model). In addition, we report the applied strain $\varepsilon_{applied}$ normalized by the critical transition strain $\varepsilon_{crit}$. From the normalized force vs. strain curve, initially in the bending energy dominated regime, $\sigma_{yy}/E \propto \tilde{\kappa}$. However, after the transitioning to the stretching energy dominated regime, $\sigma_{yy}/E$, all curves converge to a similar backbone regardless of $\tilde{\kappa}$. In short, the strain-stiffening effect  that single fiber chains exhibit under loading arises from the transition from the bending energy mode to the stretching energy mode. 

\subsubsection{Analytical Reduced Order Model Predicts Strain-Stiffening in Single Fiber Chains}
\label{sec:res_analytical_single}
\begin{figure}[ht]
    \centering
    \includegraphics[width= \textwidth]{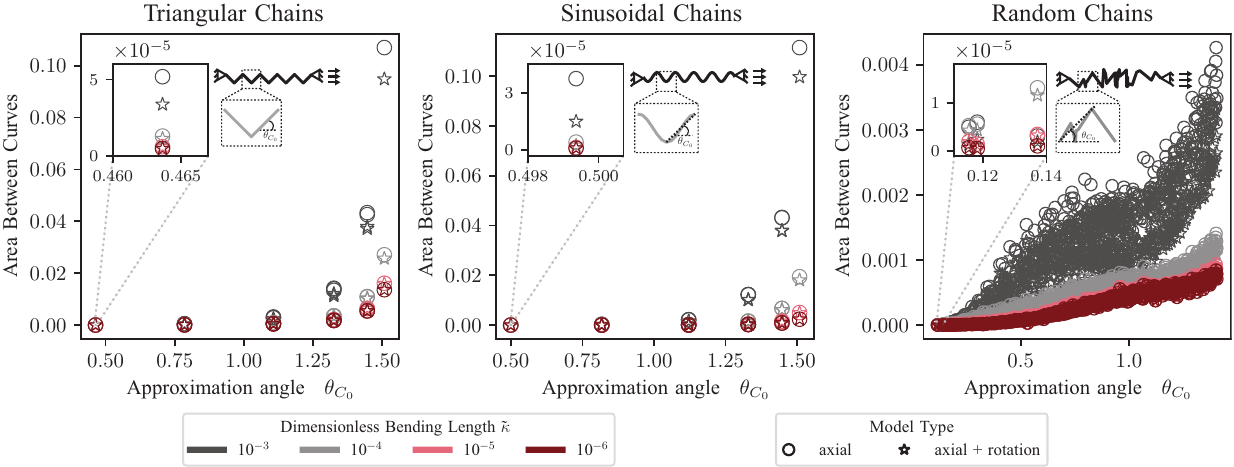}
    \caption{Absolute area between normalized force vs. strain curves from our reduced order model and FEA with different $\theta_{C_0}$ for different chain types (curves are visualized in Fig. \ref{fig:res_single_mech}). Note that for triangular chains $\theta_{C_0}$ does not have to be approximated. }
    \label{fig:res_MAPE}
\end{figure}

In this Section, we compare the results obtained from our analytical reduced model derived in Section \ref{sec:meth_rom_single} with the finite element model shown in Section \ref{sec:meth_fea}. Briefly, the analytical reduced order model combines an axial spring with a rotational spring to predict the normalized reaction force from geometric features and kinematics of the chain under uniaxial tension. In Fig. \ref{fig:res_single_mech}c, we plot the normalized force vs. strain curve obtained from FEA (the ground truth in this work), our partial reduced order model that only contains an axial spring, and our full reduced order model that combines the axial spring and rotational spring terms for our representative examples. We observe that the partial reduced order model term that only consists of the axial spring term \emph{underestimates} the normalized force, while the additional rotational term corrects this discrepancy. This discrepancy is particularly pronounced in the bending energy dominated regime where $
\sigma_{yy}/E \propto \tilde{\kappa}$. To generalize and understand the influence of the additional rotational term, we compare the partial reduced order model and the full reduced order model with the FEA solution for all single chains generated in this work by evaluating the absolute area between the curves. A small value of absolute area between the curves means that the analytical model agrees with the FEA results, and an absolute area of $0$ means that the two results coincide exactly. From Fig \ref{fig:res_MAPE}, in general, regardless of chain type, an increase in approximation angle $\theta_{C_0}$ (see Eqn. \ref{eqn:approx_angle}; intuitively a larger $\theta_{C_0}$ means a more wavy chain) leads to higher absolute area between curves (i.e., higher discrepancy between the reduced order model and FEA). And, adding the rotational spring term into our analytical model consistently leads to a smaller absolute area between curves. In the case of random fiber chains, the separation between the models with and without the rotational spring term is less clear, but the general trend of the additional rotational spring term decreasing the absolute area between curves is present except for a few rare cases. Additional details on these rare cases can be found in Appendix \ref{appendix:single_abc}. However, note that the increase of area between curves for both the partial and reduced order model for higher $\tilde{\kappa}$ might stem from an increase in magnitude of $\sigma_{yy}/E$ in the bending dominated regime. Overall, despite discrepancies between the full analytical reduced order model and FEA results, our simple analytical reduced order model that is written in terms of geometric features and kinematics can predict the mechanical behavior of single fiber chains up to reasonable accuracy. 

\subsubsection{Geometry of Single Fiber Chains Predicts the Critical Point in the Strain-Stiffening Transition}
\label{sec:res_single_phase}

\begin{figure}[ht]
    \centering
    \includegraphics[width= \textwidth]{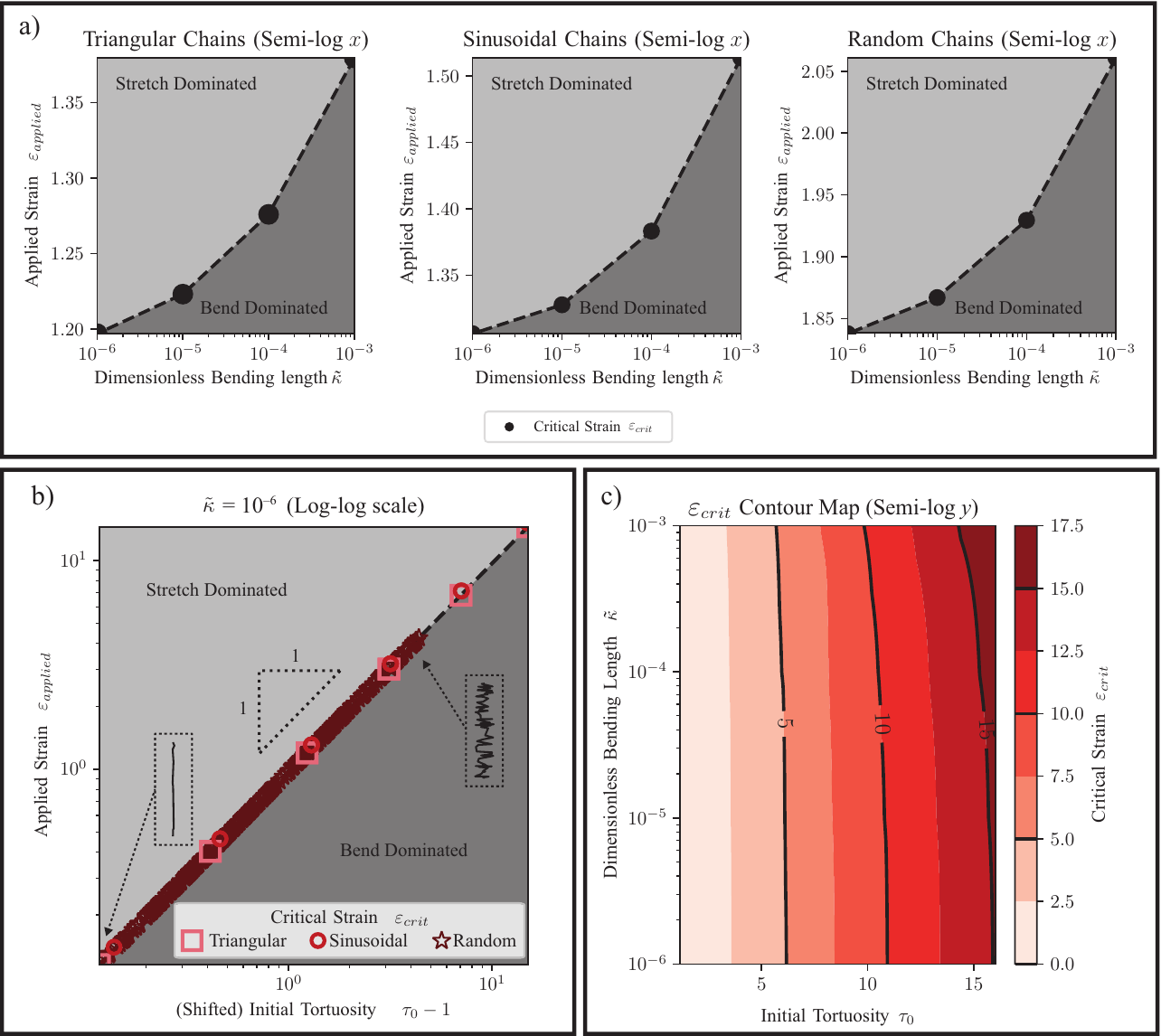}
    \caption{Relationship between critical strain $\varepsilon_{crit}$ and geometry of single fiber chains. a) Phase diagram of single fiber chains to identify the dominant deformation mode (i.e., stretching mode or bending mode) under applied strain $\varepsilon_{applied}$ with respect to the dimensionless bending length $\tilde{\kappa}$ for our representative examples. Note that the $x$-axis is in log scale. b) Phase diagram of single fiber chains to identify the dominant deformation mode (i.e., stretching mode or bending mode) under applied strain $\varepsilon_{applied}$ with respect to the initial tortuosity $\tau_0$ with dimensionless bending length $\tilde{\kappa}$ held constant at $10^{-6}$. Regardless of chain type, $(\tau_0 - 1) \sim \varepsilon_{crit}$. Note that the plot is in log-log scale. c) Contour plot of $\varepsilon_{crit}$ with respect to $\tilde{\kappa}$ and $\tau_0$. Note that the contour plot is constructed from all chain types as chain type does not affect $\varepsilon_{crit}$. Note that the $y$-axis is in log scale.}
    \label{fig:res_phase}
\end{figure}

As shown in Section \ref{sec:res_mech_single}, the mechanics of these single fiber chains is depends on the dominant deformation mode of the chain. Towards our goal of establishing structure-function relationships in fiber networks, we first investigate the strain-stiffening effect in single fiber chains. Specifically, we investigate the relationship between the critical strain transition point $\varepsilon_{crit}$ and the geometrical features of single fiber chains. First, recall that from Fig \ref{fig:res_single_mech}a, if $\varepsilon_{applied} < \varepsilon_{crit}$ (i.e., the applied strain is less than the critical transition strain), the fiber chain is in a bending dominated regime. Likewise, if $\varepsilon_{applied} > \varepsilon_{crit}$ (i.e., the applied strain is more than critical transition strain), then the fiber chain is in a stretching dominated regime. 

Figure \ref{fig:res_phase}a compares $\varepsilon_{crit}$ with $\tilde{\kappa}$ for representative examples of all chain types. From the plot, there is a clear monotonic relationship between $\varepsilon_{crit}$ and $\tilde{\kappa}$. As a result, we are able to construct a phase diagram separating the energy dominating regimes for each representative example using $\varepsilon_{crit}$ as the phase boundary between the bending dominated phase and stretching dominating phase. With this phase diagram, given the chain parameters, $\tilde{\kappa}$ and $\varepsilon_{applied}$, we can determine if the chain is in the bending energy dominated energy regime or the stretching energy dominated regime. 

Next, we look into relationship of tortuosity $\tau_0$ with respect to $\varepsilon_{crit}$. Intuitively, a larger initial tortuosity $\tau_0$ (i.e., a wavier chain) would also have a larger $\varepsilon_{crit}$ since it takes longer for the chain to unravel. In the case of a perfectly straight chain (i.e., $\tau_0 = 1$), the chain will always be in the stretching dominated regime (i.e., $\varepsilon_{crit} = 0$). With this in mind, in Fig \ref{fig:res_phase}b, we plot $\varepsilon_{crit}$ with respect to $\tau_0-1$ on a log-log scale for all chain types. We use $\tau_0-1$ to ensure that the log-log plot is bounded since $\varepsilon_{crit} \rightarrow 0$ as $\tau_0 \rightarrow 1$. From the results of Fig. \ref{fig:res_phase}b, $(\tau_0 - 1) \sim \varepsilon_{crit}$, regardless of chain type, there is a linear relationship between $(\tau_0-1)$ and $\varepsilon_{crit}$ (and subsequently $\tau_0$ and $\varepsilon_{crit}$) since there is a slope of $1$ in the log-log plot. Again, following the same logic as Fig \ref{fig:res_phase}a, due to the clear linear relationship between $\tau_0-1$ and $\varepsilon_{crit}$, we are able to construct a phase diagram that determines the dominant deformation mode that the single fiber chain is in. With clearly established relationships between geometrical features $\tilde{\kappa}$ and $\tau_0$ with $\varepsilon_{crit}$ that is agnostic to the type of chain, we are able effectively summarize the structure-function relationship of single fiber chains by constructing a contour plot that predicts $\varepsilon_{crit}$ just from geometrical features in Fig. \ref{fig:res_phase}c. On the same note, using the same arguments as above, this contour plot of $\varepsilon_{crit}$ can be interpreted as a visualization of the phase boundary between the bending energy dominated regime and the stretching energy dominated regime. In conclusion, we demonstrated that the mechanical response of single fiber chains (i.e., function), which depends on the dominating deformation mode,  can be determined purely through its kinematics and geometric features (i.e., structure).

\subsection{Mechanics of Random Fiber Networks}
\label{sec:res_mech_fibers}
\begin{figure}[ht]
    \centering
    \includegraphics[width= \textwidth]{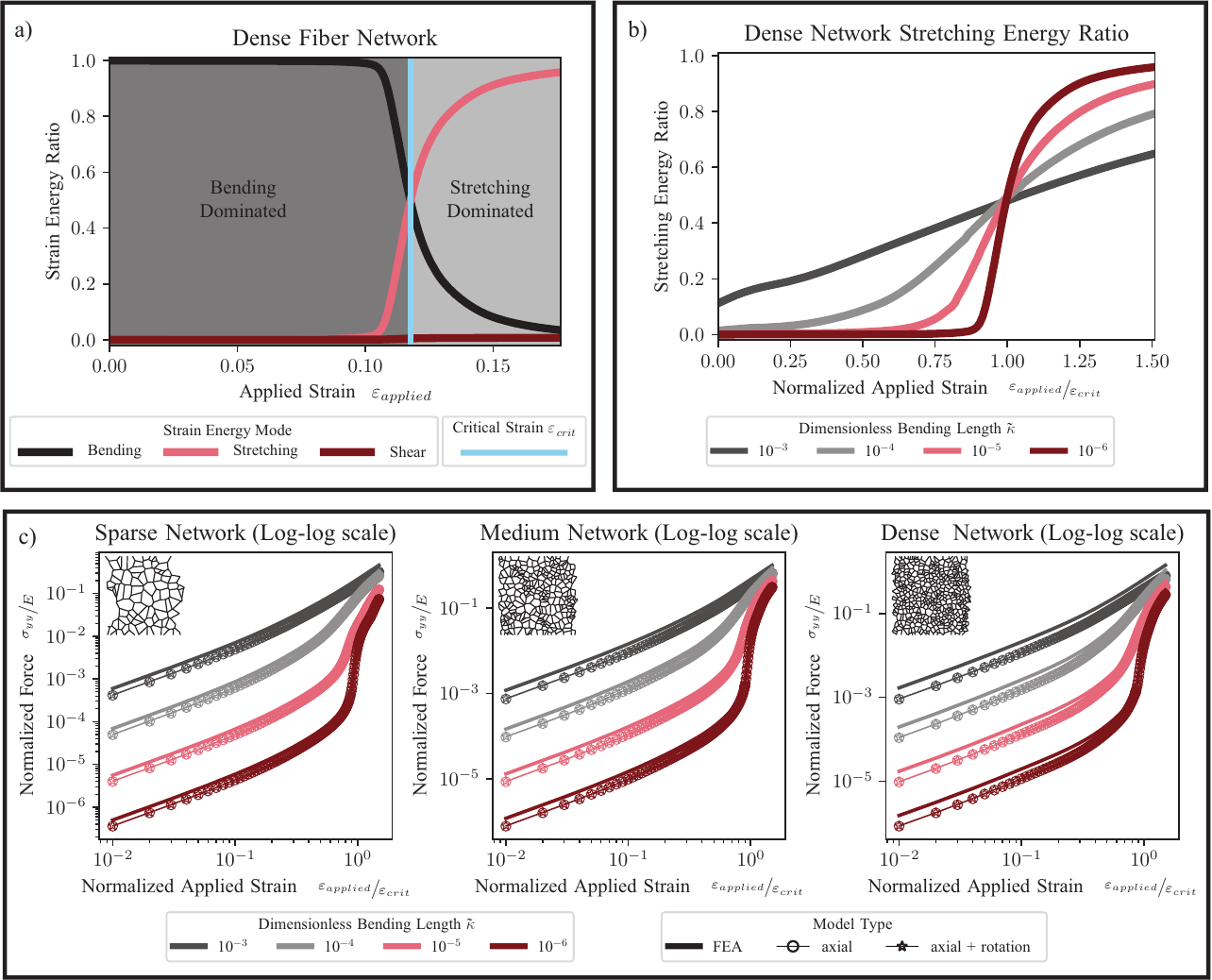}
    \caption{Representative examples to demonstrate the mechanics of random fiber networks. a) Strain energy distribution of random fiber network under uniaxial tension. The representative example shown is from a dense network with $\tilde{\kappa}=10^{-6}$.  b) Comparisons of stretching energy distribution for different $\tilde{\kappa}$.  c) Comparisons between the performance of our analytical reduced order model and FEA results on predicting the dimensionless reaction force $\sigma_{yy}/E$ for all $3$ types of single fiber chains. Note that the plots are in log-log scale.}
    \label{fig:res_fiber_mech}
\end{figure}

With the structure-function relationships of the single fiber chains established, we now extend this analysis to random fiber networks. Here, we start by making direct comparisons between the mechanics of random fiber networks and the mechanics of single fiber chains. In Fig. \ref{fig:res_fiber_mech}a, we look at the change in the strain energy ratio (see Eqn. \ref{eqn:enery_ratio}) under uniaxial loading of a representative random fiber network ($n=500$, $\tilde{\kappa} = 10^{-6}$). \changes{Despite having a more complicated structure, random fiber networks exhibit the same bending to stretching energy transition as single fiber chains. Thus, we also define the critical strain transition point as $\varepsilon_{crit}$ for fiber networks.} Similarly, the shear deformation mode is significantly lower than the bending and stretching energy modes. \newchanges{From Fig. \ref{fig:res_fiber_mech}b, we observe that the transition between dominant energy modes depends on $\tilde{\kappa}$ where a random fiber network with a larger $\tilde{\kappa}$ (i.e., thicker fibers) will have a more gradual transition between the energy modes compared to a random fiber network with a smaller $\tilde{\kappa}$ (i.e., thinner fibers).} \changes{As a result of the random fiber network transitioning from a bending dominant regime to a stretching dominated regime, in Fig \ref{fig:res_fiber_mech}c we can observe a strain stiffening effect in the normalized force $\sigma_{yy}/E$ vs. strain $\varepsilon_{applied}/\varepsilon
_{crit}$ curves for our representative examples ($n=100$ for ``sparse'' networks; $n=300$ for ``medium'' networks; $n=500$ for ``dense'' networks), which is in line with other work in literature \citep{parvez2023stiffening, sharma2016strain}.} Similar to single fiber chains, initially, $\sigma_{yy}/E \propto \tilde{\kappa}$ in the bending energy dominated regime before converging to a similar backbone in the stretching dominated regime independent of $\tilde{\kappa}$. In summary, the mechanical response of random fiber networks is analogous to the mechanical response of single fiber chains. However, as the following Sections will demonstrate, the mechanism behind this strain-stiffening effect in random fiber networks is more complex than that of when compared to single fiber chains. 

\subsubsection{Shortest Paths Predict the Mechanics of Random Fiber Networks}
\label{sec:res_fiber_abc}

\begin{figure}[ht]
    \centering
    \includegraphics[width= \textwidth]{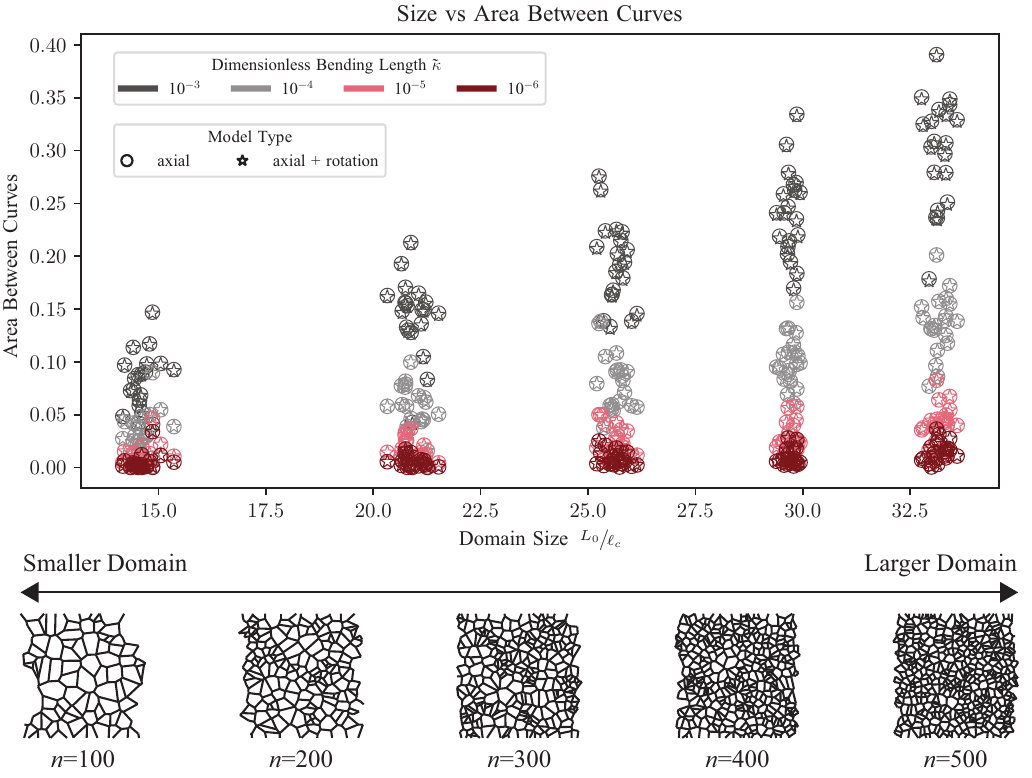}
    \caption{Absolute area between force-displacement curves from our reduced order model and FEA results (curves are visualized in Fig. \ref{fig:res_fiber_mech}). All the data points shown are for all fiber networks with initial $n \in \{ 100,200,300,400,500 \}$ random Voronoi seeds.}
    \label{fig:res_fiber_abc}
\end{figure}

In Section \ref{sec:res_analytical_single}, we demonstrated that the analytical reduced order model derived in Section \ref{sec:meth_rom} is able to predict the mechanics of single fiber chains relatively well. In this Section, we investigate the performance of our reduced order model for random fiber networks. In brief, to translate our analytical reduced order model of single fiber chains to random fiber networks, we first compute the distance-weighted shortest paths between the constrained boundaries and interpret the shortest paths as single fiber chains in parallel (see Section \ref{sec:meth_rom_network}). In Fig. \ref{fig:res_fiber_mech}c, we plot the normalized force $\sigma_{yy}/E$ vs. strain $\varepsilon_{applied}/\varepsilon
_{crit}$ obtained from FEA, our partial reduced order model (i.e., axial spring only), and full reduced order model (i.e., axial and rotational spring) for our representative examples. Similar to single fiber chains, the partial reduced order model is able to predict the normalized force relatively well, but still underestimates the normalized force when compared to FEA results. This discrepancy between the partial reduced order model and FEA result is more pronounced in the bending energy dominated regime where $\sigma_{yy}/E \propto \tilde{\kappa}$. However, in contrast to single fiber chains, the additional rotational spring term does not correct for this discrepancy. To generalize this notion for all types of random fiber networks simulated, we compare the absolute area between curves between the partial and full reduced order model with FEA solution for different domain size $L_0/\ell_C$ (see Fig. \ref{fig:res_fiber_abc}). Note that in this work, to investigate size effects, the domain size is varied by prescribing a constant $L_0 = 10,000$ and manipulating $\ell_C$ by changing the number of Voronoi seeds (see bottom of Fig. \ref{fig:res_fiber_abc}). Our range of domain size was selected to ultimately meet the recommended size such that size effects are negligible \citep{merson2020size,parvez2024methodological,shahsavari2013size}. Overall, consistent with our observations from Fig. \ref{fig:res_fiber_mech}c, the additional rotational spring term does not correct for this discrepancy. And, similar to single fiber chains, there is an increase in absolute area between curves as $\tilde{\kappa}$ increases, which may be a direct result of the relationships $\sigma_{yy}/E \propto \tilde{\kappa}$ in the bending dominated regime. Additionally, there seems to be a larger range of error (i.e., absolute area between curves) as the domain size increases, which seems to suggest that our reduced order model works better for smaller domains. In summary, we have shown that our simple analytical reduced order model constructed by interpreting the shortest paths in the networks as springs in parallel can capture the mechanics of random fiber networks fairly well (worse case mean area between curves $ \approx 0.192$ for random fiber networks with $n=500$), but the additional rotational spring no longer facilitates in correcting the linear spring term like in the case of single fiber chains.

\subsubsection{Emergence of Force Chains and Support Networks in Random Fiber Networks}
\label{sec:res_forcechain}

\begin{figure}[p]
    \centering
    \includegraphics[width= \textwidth]{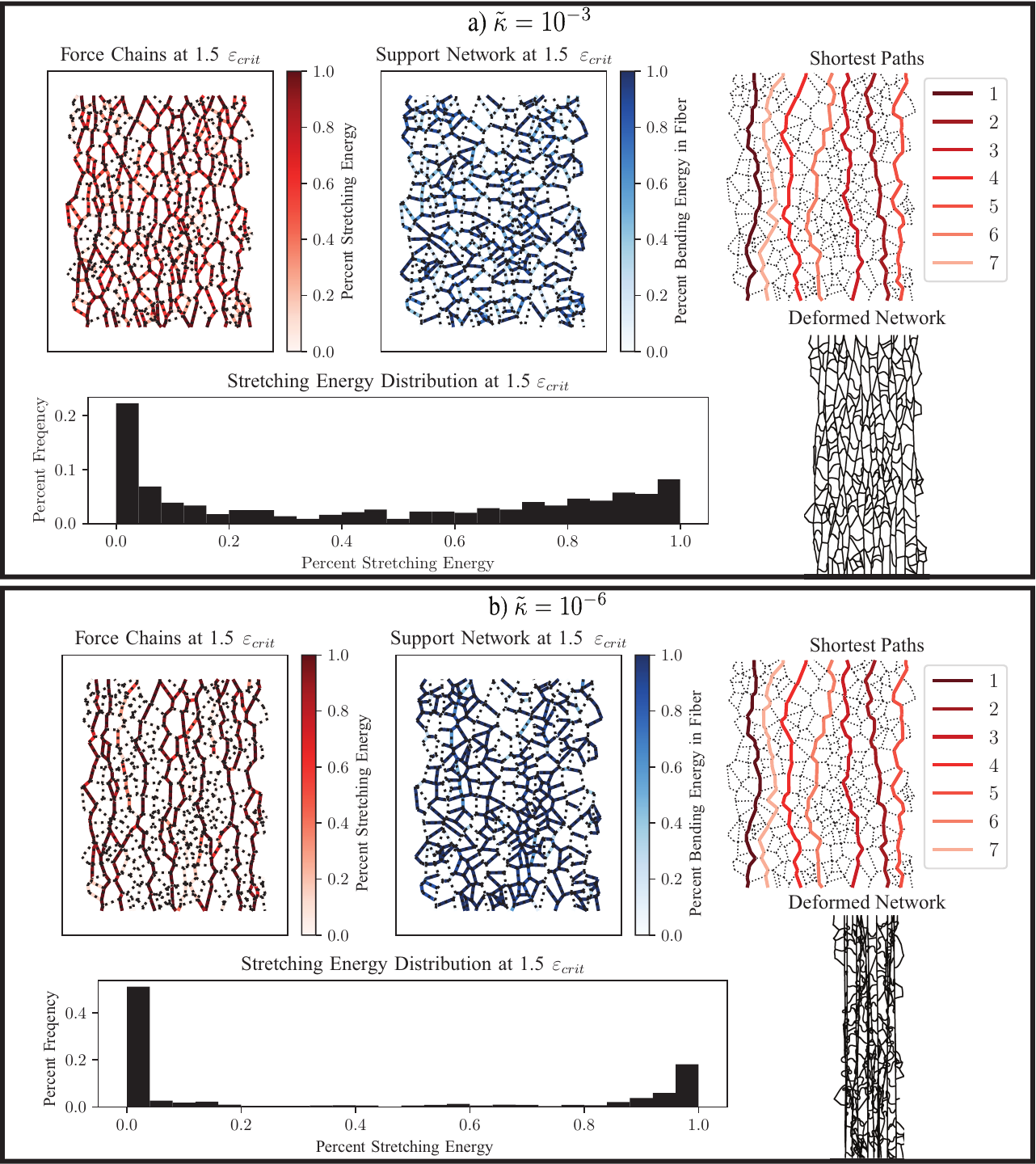}
    \caption{Representative example of medium fiber networks of the percent energy distribution in each fiber for applied strain $\varepsilon_{applied} = 1.5 \varepsilon_{crit}$. Note that for the plots, each energy component and energy percent is computed individually for each fiber (i.e., a stretching energy percent of $1.0$ means that the individual fiber is only stretching and not bending). a) Energy distribution in each fiber, histogram of energy carried in each fiber, and distance-weighted shortest paths and  deformed configuration for $\tilde{\kappa} = 10^{-3}$ (i.e., thicker fibers). b) Energy distribution in each fiber, histogram of energy carried in each fiber, deformed configuration, and distance-weighted shortest paths for $\tilde{\kappa} = 10^{-6}$ (i.e., more slender fibers). }
    \label{fig:res_fiber_dist}
\end{figure}

\begin{figure}[p]
    \centering
    \includegraphics[width= \textwidth]{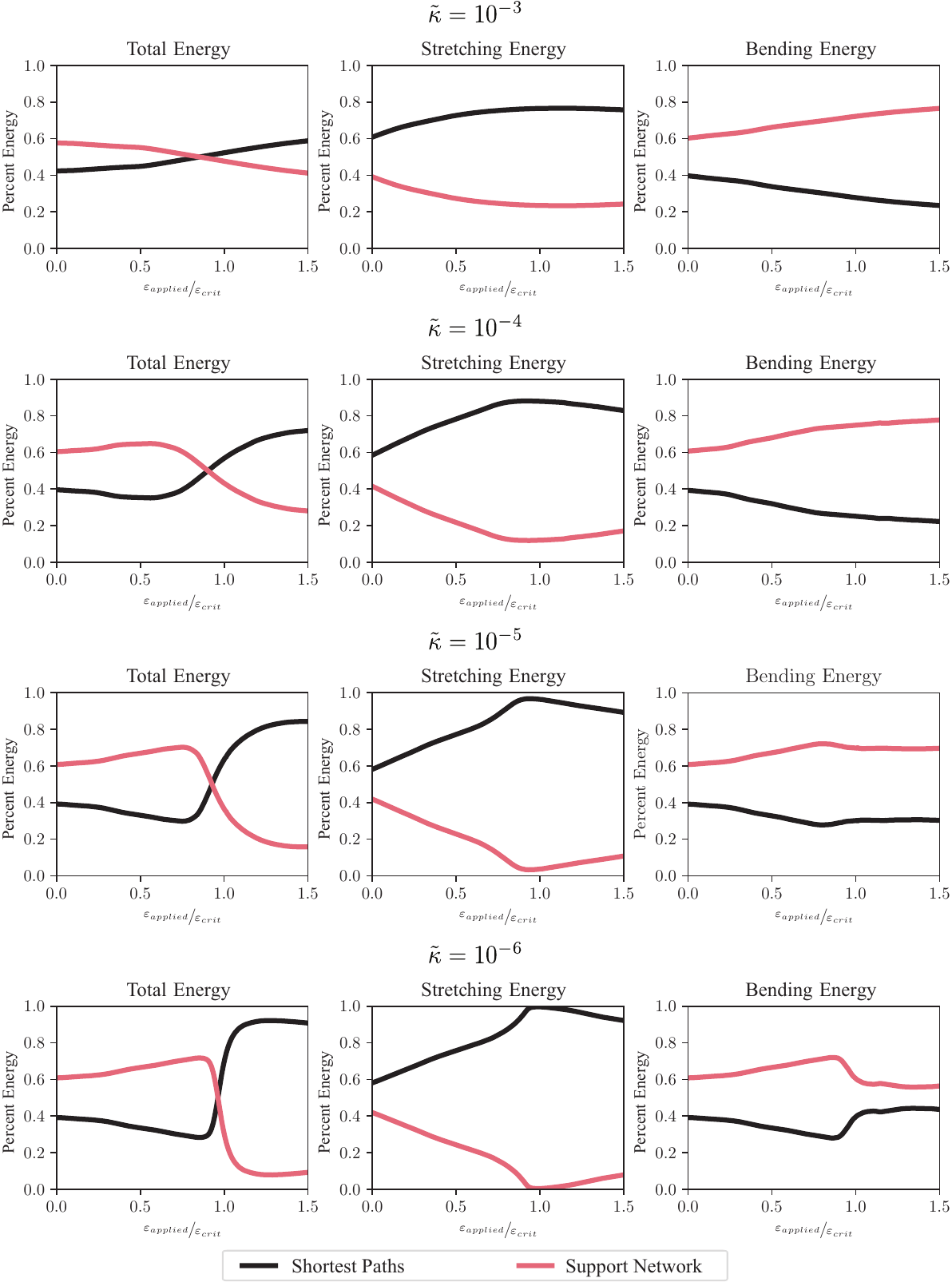}
    \caption{Percent total energy, percent stretching energy, and percent bending energy in the random fiber network partitioned by shortest paths (black line) and support network (pink line) for $\tilde{\kappa} = \{10^{-3}, 10^{-4}, 10^{-5}, 10^{-6} \}$. }
    \label{fig:res_short_sup}
\end{figure}

While shortest paths are able to predict the mechanics of random fiber networks relatively well, in this Section, we explore the mechanistic structure-function relationship of random fiber network by investigating the connection between the distance-weighted shortest paths, force chains, and mechanical behavior. In brief, force chains are the load paths that emerge in random fiber networks under applied load. Specifically, towards connecting the geometric structure of random fiber networks to their mechanical function, we investigate the feasibility of our shortest paths algorithm proposed in Section \ref{sec:meth_rom_network} as an alternative method to identify force chains under uniaxial loading without needing additional information about the force distribution in random fiber networks. And, we examine the mechanical contribution of the remainder of the network excluding the shortest paths, termed the ``support network''. To this end, we first investigate the local strain energy distributions of a representative random fiber network with medium density (i.e., $n=300$) at an $\varepsilon_{applied} = 1.5 \varepsilon_{crit}$ (i.e., the random fiber network is in stretching dominated regime). The local strain energy ratio is defined similarly as Eqn. \ref{eqn:enery_ratio}, but with the energy ratio computed locally for each fiber instead of globally for the whole chain/network. In Fig. \ref{fig:res_fiber_dist}a and Fig. \ref{fig:res_fiber_dist}b we visualize the local energy ratio, local energy distribution, and the deformed configuration for the same representative fiber network with $\tilde{\kappa} = 10^{-3}$ (i.e., thicker fibers) and $\tilde{\kappa} = 10^{-6}$ (i.e., thinner fibers) respectively. In general, the fibers that are stretching dominated are aligned parallel to the direction of loading, while the fibers that are bending dominated are perpendicular to the direction of loading (here the fiber networks are loaded in the vertical direction; see Fig. \ref{fig:meth_examples} for visualization of the boundary conditions). Note that the occurrence of fibers in stretching and bending dominating modes exhibiting dependence on the loading direction has also been observed in a different work on the stiffening effect of random fiber networks \citep{parvez2023stiffening}. In addition to the dependence on the loading direction, another difference between the fibers in the bending dominated regime and fibers in the stretching dominated regime is their spatial distribution. Fibers that are stretching dominated are connected and form force chains between the two constrained boundaries, while the fibers that are bending dominated are more amorphous. \changes{Qualitatively, the fibers that are stretching dominated look similar to our distance-weighted shortest path, albeit the actual force chains are often more complex than our distance-weighted shortest paths.} Comparing the strain energy distributions for our representative example of random fiber networks with $\tilde{\kappa} = 10^{-3}$ (Fig. \ref{fig:res_fiber_dist}a; network with thicker fibers) with the same representative example with $\tilde{\kappa} = 10^{-6}$ (Fig. \ref{fig:res_fiber_dist}b; network with thinner fibers), the strain energy distribution for the fiber network with $\tilde{\kappa} = 10^{-3}$ is more uniform, with more fibers not completely stretching dominated or bending dominated. \changes{As a result, the fiber network with $\tilde{\kappa} = 10^{-3}$ have a more uniform energy distribution. Visually, the force chains for the random fiber network with $\tilde{\kappa} = 10^{-3}$ exhibit more branching (i.e., the chains ``split'' into multiple paths) and coalescing (i.e., multiple paths ``merge'' into one path)}. The more homogeneous nature of the strain energy also affects the deformed configuration, where the deformed network for $\tilde{\kappa} = 10^{-6}$ has a smaller Poisson's effect. Specifically, this Poisson's effect emerges from fibers in the  support network bending, which in turn pulls the network together. From the stretching energy histogram, since the network $\tilde{\kappa} = 10^{-3}$ has less fibers that are strictly bending, the Poisson's effect is less pronounced. Additionally, the deformed force chains are straighter in the case of $\tilde{\kappa}=10^{-6}$, again since more fibers are strictly stretching dominated.

To ensure that the shortest paths we selected via the algorithm presented in Section \ref{sec:meth_rom_network} actually correspond to force chains, we look at the percent energy ratio of the shortest paths and the support network for the same representative fiber network with $\tilde{\kappa} = \{ 10^{-6},10^{-5},10^{-4},10^{-3} \}$ in Fig. \ref{fig:res_short_sup}. More specifically, we look at the energy partition
$\Pi^{p}_{[\dots]}/\Pi_{[\dots]}$ where $p$ denotes the subsets of fibers of interest (i.e., shortest paths or support network), and $\Pi_{[\dots]}$ is the energy we are investigating (i.e., bending or stretching energy). The shortest paths should carry a majority of the stretching energy in the fiber network if they correspond to the force chains, since stretching energy translates to axial force. 

Starting from the right column in Fig. \ref{fig:res_short_sup}, in general, the support network contains most of the total strain energy in the bending energy dominated phase, while the shortest paths contain most of the total strain energy in the stretching dominated phase. This effect is more pronounced for the random fiber network with $\tilde{\kappa} = 10^{-6}$, which matches with our observations from Fig. \ref{fig:res_fiber_dist}, where the fiber network with $\tilde{\kappa} = 10^{-6}$ has fibers that are either completely bending dominated or completely stretching dominated. On the other hand, in the case of total stretching energy (middle column) and total bending energy (left column), there is no transition between dominant energy modes. In other words, the shortest paths always contain the majority of the stretching energy, while the support network always contains the majority of the bending energy.  Additionally, in the case of stretching energy, there is a clear separation in energy distribution where the shortest paths contain almost all the stretching energy after the strain-stiffening transition, especially for the case where $\tilde{\kappa} = 10^{-6}$, while the discrepancy in bending energy is not as apparent. \changes{However, as mentioned above, our method of predicting force chains with distance-weighted graph shortest path is able to capture branching and coalescing of force chains. Being able to capture these complex force chain interactions can lead to a better understanding of structure-property relationships of random fiber networks.}

\changes{Overall, despite only capturing basic force chain interactions, we are able to show that our shortest paths obtained from random fiber network geometric structure alone \textit{without any force information} carry the majority of the stretching energy.} Additionally, we observed that the support network performs the opposite function to our shortest paths by containing most of the bending energy in the network. As such, through our shortest paths algorithm outlined in Section \ref{sec:meth_rom_network}, we are able to identify two subsets of fibers that perform different mechanical functions just from the geometric structure of the random fiber network. As a side note, the support network containing most of the bending energy might be the main reason that the additional rotational spring term does not help improve the partial reduced order model. Trying to incorporate the support network into the analytical model could be an interesting future research direction that is outside the scope of this work.  

\subsubsection{Shortest Paths Recruit Sequentially During Loading}
\label{sec:res_recruitment}

\begin{figure}[p]
    \centering
    \includegraphics[width= \textwidth]{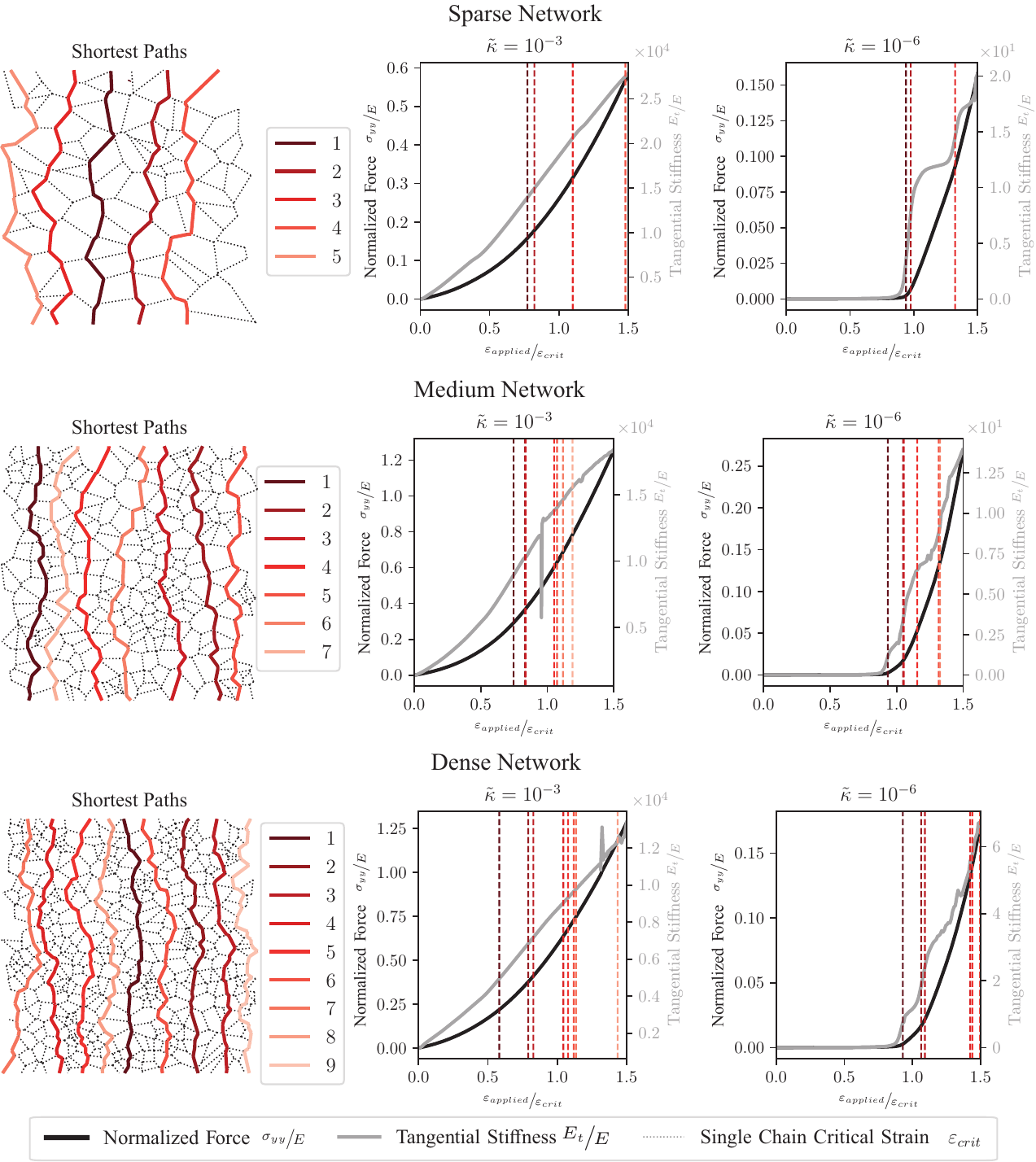}
    \caption{Demonstration of fiber recruitment for representative examples of random fiber networks. The first column illustrates the shortest paths in the network, with the color denoting the shortest path number. The middle column and right column contains a dual axis plot visualizing normalized force $\sigma_{yy}/E$ (black line) and tangential stiffness $E_t/E$ (gray line). The dotted lines denote the $\varepsilon_{crit}$ of each shortest paths with the line color denoting the same shortest path number as visualized in the left column.}
    \label{fig:res_fiber_recruit}
\end{figure}

To further support the idea that the shortest paths defined in Section \ref{sec:meth_rom_network} dictate the mechanical function of random fiber networks, we examine the role of the shortest paths in strain-stiffening. More specifically, we interpret our shortest paths as individual fiber chains that sequentially recruit during loading. This fiber recruitment behavior is essential in the nonlinear mechanics of soft fibrous tissue \citep{jan2018collagen, mattson2019contributions}. 
To perform this analysis, we treat these fiber networks as a simpler mechanical system. Specifically, we ignore the support network and consider only the shortest paths as a collection of springs in parallel. 
From Section \ref{sec:meth_rom_network}, note that for the set of shortest paths $\Phi = \{ \phi^1, \phi^2, \dots , \phi^k \}$, the corresponding total path distance is $D(\phi^1) \leq D(\phi^2) \dots \leq D(\phi^k)$ (i.e., the shortest paths are arranged from the smallest total path distance to the largest total path distance). We can individually compute the initial tortuosities of each shortest path as:

\begin{equation}
\tau_0^k = \frac{D(\phi^k)}{L_0} \, .
\end{equation}

Note that the superscript $k$ here denotes that the variable is individually computed from our $k^{th}$ shortest path. From our observations in Section \ref{sec:res_single_phase}, by ignoring the support network, the individual $\varepsilon_{crit}^k$ of each shortest path can be inferred from the geometric features $\tilde{\kappa}$ and $\tau_0^k$. As such, as $\varepsilon_{applied}$ is applied to the system of shortest paths, initially, each path will be in the bending dominated regime. As $\varepsilon_{applied} \rightarrow \varepsilon_{crit}^k$, the $k^{th}$ shortest path will transition from the bending dominated regime to the stretching dominated regime (i.e., the path is ``recruited''), which in turn will result in a sequential strain-stiffening effect. In Fig. \ref{fig:res_fiber_recruit} we investigate the role of our shortest paths in fiber recruitment by plotting $\varepsilon_{crit}^k$ of each individual path with respect to normalized force $\sigma_{yy}/E$ and normalized tangential stiffness $E_t/E$, with the tangential stiffness $E_t$ being defined as:

\begin{equation}
E_t = \frac{d \sigma_{yy}}{d \varepsilon_{applied}}.
\end{equation}

Here, differentiation for $E_t$ is done numerically using NumPy \citep{harris2020array} with a central finite difference scheme. A forward/backward scheme is used on the boundaries as appropriate. From Fig. \ref{sec:res_single_phase}, we observe that in the case where the random fiber networks have $\tilde{\kappa} = 10^{-6}$ (i.e., thinner fibers), regardless of domain size, there is a sharp increase in $E_t$ every time a new path is recruited. However, in the case where the fiber networks have $\tilde{\kappa} = 10^{-3}$ (i.e., thicker fibers), the fiber recruitment effect is less clear because $E_t$ increases relatively linearly, aside from what may be numerical artifacts from differentiation. The slower gradual increase in stiffness could be due to the slower change in energy transition for thicker single chains (see Fig.\ref{fig:res_single_mech}b) and networks with thicker fibers (see Fig. \ref{fig:res_fiber_mech}b), which in turn smooths out the jumps in $E_t$. Additionally, the more gradual change in stiffness for fiber networks with $\tilde{\kappa} = 10^{-3}$ might also stem from our observations in the previous sections where the support network plays a more important mechanical function in fiber networks with thicker fibers due to a larger bending stiffness. This is in contrast to our simple shortest paths only system, which can be interpreted as the case where $\tilde{\kappa} \rightarrow 0$ for the support network. In short, we are able to observe our shortest paths recruiting sequentially as the fiber network stiffens, with the sequential stiffening effect being more pronounced for networks with thinner fibers.

\subsubsection{Prediction of Critical Strain-Stiffening Point From Shortest Paths}
\label{sec:res_fiber_phase}

\begin{figure}[p]
    \centering
    \includegraphics[width= \textwidth]{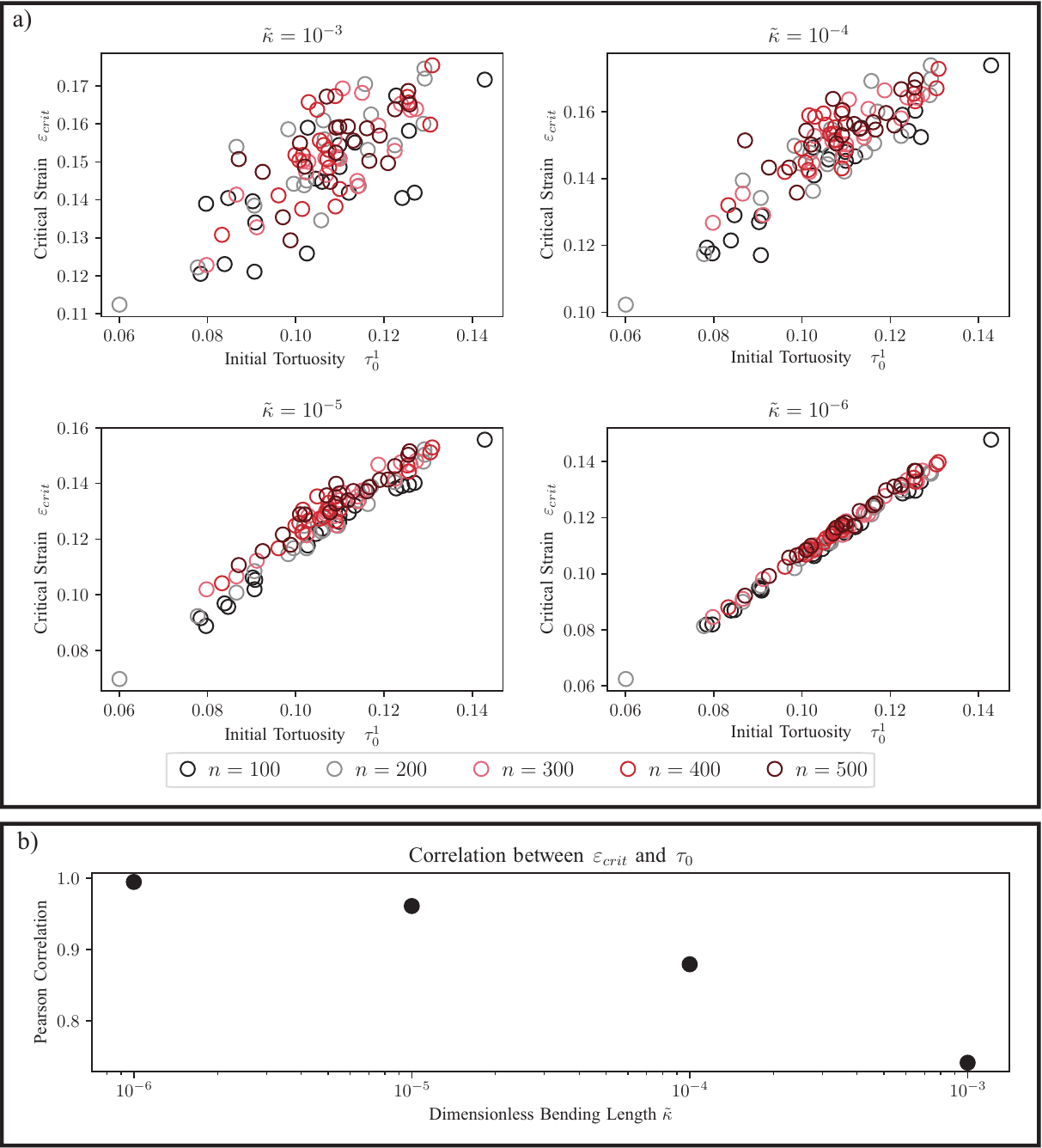}
    \caption{Relationship of critical strain $\varepsilon_{crit}$ and geometric parameters for random fiber networks. a) Relationship between initial tortuosity $\tau_0$ and critical strain $\varepsilon_{crit}$ for different dimensionless bending length $\tilde{\kappa}$ with fiber network with $n \in \{ 100,200,300,400,500 \}$ initial Voronoi seeds. b) Quantifies the linear relationship between $\varepsilon_{crit}$ and $\tau_0$ for all random fiber networks through Pearson Correlation coefficient.}
    \label{fig:res_fiber_phase}
\end{figure}

Now that we have established that the shortest paths do have a mechanical function in random fiber networks, to further infer mechanical function from geometric structure, we investigate the feasibility of using the shortest paths and kinematics to predict the energy dominating phase that the fiber network is currently under (i.e., a result analogous to Section \ref{sec:res_single_phase}). More intuitively, if we again ignore the support network (i.e., only leaving the shortest paths as springs in parallel), the system will be in the bending dominated regime until the first shortest path is recruited (i.e., least wavy shortest path; the first shortest path by definition). This observation is consistent with Fig. \ref{fig:res_fiber_recruit}, where the first shortest path always recruits (i.e., $\varepsilon_{applied} > \varepsilon_{crit}^1$ ) before the fiber network globally transitions to a stretching dominated state (i.e., $\varepsilon_{applied} > \varepsilon_{crit}$). As such, we use $\tau_0^1$ (i.e., tortuosity of the first shortest path) as our geometric parameter to infer the $\varepsilon_{crit}$ for our random fiber network. In Fig. \ref{fig:res_fiber_phase}a we plot $\varepsilon_{crit}$ vs. $\tau_0$ for all $\tilde{\kappa}$ and for all domain sizes (i.e., $n=\{ 100,200,300,400 \}$). We observe that regardless of domain size $L_0/\ell_C$, there is a linear relationship between $\tau_0^1$ and $\varepsilon_{crit}$ that grows stronger for smaller $\tilde{\kappa}$. Specifically, as $\tilde{\kappa} \rightarrow 10^{-6}$, the plots converge towards a linear relationship like the one realized in Fig. \ref{fig:res_phase}b. The dependence of the strength of correlation is characterized in Fig. \ref{fig:res_fiber_phase}b where we plotted the Pearson's Correlation Coefficient between $\tilde{\kappa}$ and $\tau_0^1$. In brief, the Pearson's correlation ranges from $[-1,1]$, and characterizes the strength of linear relationship between two variables, where a Pearson's Correlation closer to $\pm 1$ means a stronger linear relationship. Fig. \ref{fig:res_fiber_phase}b reconfirms our observation from Fig. \ref{fig:res_fiber_phase}b where a smaller $\tilde{\kappa}$ has a higher linear dependence between $\tau_0^1$ and $\varepsilon_{crit}$. This observation is compatible with our intuition with the simple system that ignores the support network (i.e., the case where $\tilde{\kappa} \rightarrow 0$ for the support networks). Additionally, the stronger correlation between $\tau_0^1$ and $\varepsilon_{crit}$ for smaller $\tilde{\kappa}$ can also be observed in Fig. \ref{fig:res_fiber_recruit}, where the first shortest path is recruited close to $\varepsilon_{crit}$ for fiber networks with $\tilde{\kappa} = 10^{-6}$ when compared to fiber networks with $\tilde{
\kappa} = 10^{-3}$. 

Beyond the relationship between $\tau_0^1$ and $\varepsilon_{crit}$, we also examine the relationship between $ \tilde{\kappa}$ and $\varepsilon_{crit}$ in Appendix \ref{appendix:fiber_phase}. Overall, while there is a trend between $\varepsilon_{crit}$ and geometric features of the random fiber network, the relationship is not clear enough such that a phase diagram and a summary contour plot can be constructed in a similar manner as Section \ref{sec:res_single_phase}. Future work can be performed to refine the prediction of energy regimes of random fiber networks using this work as a stepping stone. 

\subsubsection{Network Sensitivity to Fiber Ablation}
\label{sec:res_ablation}

\begin{figure}[ht]
    \centering
    \includegraphics[width= \textwidth]{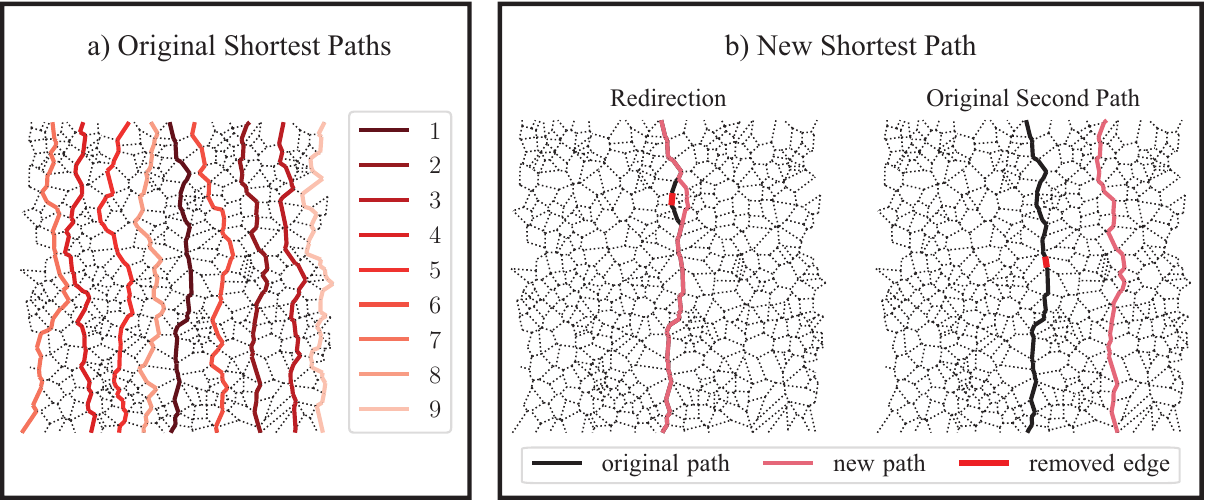}
    \caption{\changes{New shortest path of random fiber networks after ablation. a) Initial shortest paths of our representative fiber network. b) Types of new shortest paths after ablating one fiber in the first shortest path. The first type of new shortest path consists of fibers mostly from the original first shortest path, but with path redirection that circumvents the ablated fiber. The second type of new shortest path is when the new shortest path is completely different from the original first shortest path. In this case, the new first shortest path from the ablated network is the second shortest path from the original network.}}
    \label{fig:res_path_redirect}
\end{figure}

\begin{figure}[p]
    \centering
    \includegraphics[width= \textwidth]{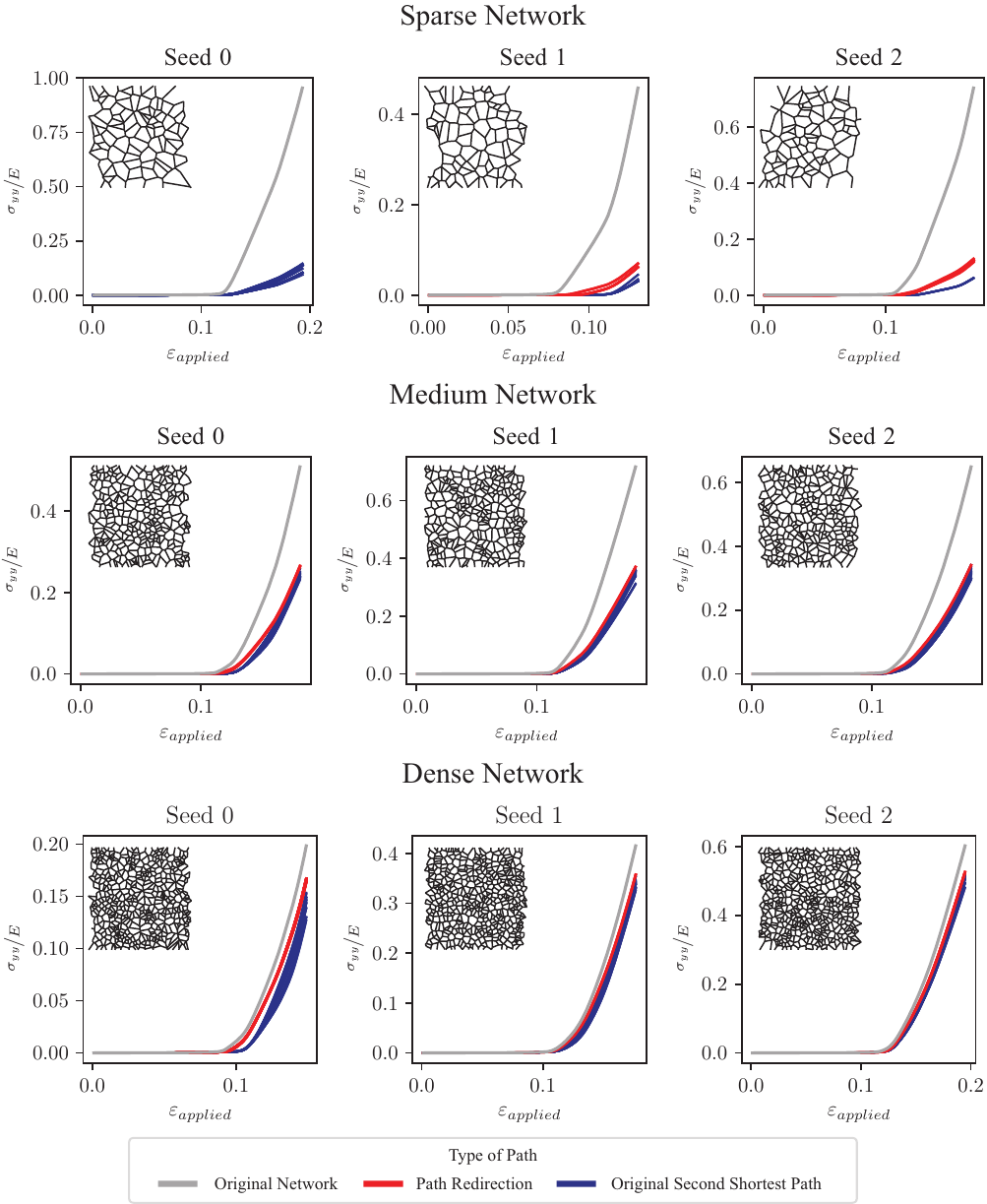}
    \caption{\changes{Normalized force vs. applied strain for representative random fiber networks with different domain sizes. Each plot contains all possible single fiber ablations from the original network's first shortest path. The gray lines represent the normalized force vs. applied strain from the original network, while the red solid line and the blue solid line denotes the normalized force $\sigma_{yy}/E$ vs. applied strain curve $\varepsilon_{applied}$ for networks with new shortest path from redirection and new shortest path from the original network's second shortest path respectively. Note that for the case of sparse network with seed $0$, all the new paths are from the original network's second shortest path.}}
    \label{fig:res_ablation_fd}
\end{figure}

\changes{
In previous Sections, we used Voronoi diagrams as the geometric domain for our random fiber networks. In many cases, the reference configuration of the geometric domain can change in biological systems due to remodeling \citep{ambrosi2011perspectives}, catch/slip bonds \citep{rakshit2012ideal} or tissue damage/healing\citep{das2021extracellular}. For engineered architected materials, mechanical properties can be altered by pruning bonds \citep{reid2018auxetic}. In both the case of biological systems and engineered materials, global change in mechanical behavior is induced from local changes in geometric structure. As a step towards understanding the complex interplay between local changes in geometric structure and their effect on global mechanical behavior, we investigate the sensitivity of mechanical behavior to fiber ablations in the reference configuration. Specifically, we sequentially ablate one of the fibers in the first shortest path, and set the ablated fiber network as the new reference configuration. This process is repeated on the original fiber network for all fibers on the first shortest paths to generate the set of ablated fiber network used in this study.}

\changes{Before looking into the mechanics of ablated fiber networks, we first investigate the geometric changes to the first shortest path between the original fiber network and the ablated fiber network. We visualize a representative example for a Voronoi network with $n=500$ initial seeds in Fig. \ref{fig:res_path_redirect}. First, all the $k^{th}$ shortest paths $\phi^k$ in the original fiber network are visualized in Fig. \ref{fig:res_path_redirect}a. Once a fiber from the first shortest path $\phi^1_{og}$ of the original fiber network is ablated, the new shortest path $\phi^1_{new}$ that form can be either be of two cases: 1) a redirected path where most of the original path $\phi^1_{og}$ is still intact, with the path redirection appearing near the vicinity of the ablated fiber (see left of Fig. \ref{fig:res_path_redirect}b), or 2) a completely different path from $\phi^1_{og}$, where the second shortest path of the original fiber network will now become the first shortest path of the ablated fiber network (see right of Fig. \ref{fig:res_path_redirect}b). As a result, we can identify the type of new shortest path by simply comparing at the ratio between the path lengths between the shortest path from the ablated fiber network $D(\phi^1_{new})$ and the \emph{second} shortest path of the original random fiber network $D(\phi^2_{og})$. If $D(\phi^1_{new}) < D(\phi^2_{og})$, then path redirection occurred; on the other hand, if $D(\phi^1_{new}) = D(\phi^2_{og})$, then the second shortest path of the original fiber network now becomes the first shortest path of the ablated fiber network. Note that the case $D(\phi^1_{new}) > D(\phi^2_{og})$ can not be observed by our definition of shortest paths.}  

\changes{Next, we look at the normalized force ${\sigma_{yy}}/{E}$ vs. applied strain $\varepsilon_{applied}$ curves to study how ablating fibers in $\phi^1_{og}$ affects the mechanical behavior of the fiber networks. Note for visual comparison we stop the simulations when $\varepsilon_{applied} = \varepsilon_{crit}$ of the original random fiber networks. Additionally, we investigate the effect that the different type of new path has on the mechanical response of random fiber networks. In Fig. \ref{fig:res_ablation_fd} we plot the $\sigma_{yy}/{E}$ vs. $\varepsilon_{applied}$ curve for the original fiber network (gray line), ablated fiber network with redirect paths (red line), and ablated fiber networks with completely new first shortest path (blue line) for sparse ($n=100$), medium ($n=300$) and dense ($n=500$) representative networks. Note that identification the type of new path is done by comparing path lengths as described above. In general, removing a fiber from the shortest path decreases the overall stiffness (i.e., slope of the $\sigma_{yy}/{E}$ vs. $\varepsilon_{crit}$ curve). In addition, qualitatively, the critical strain transition point $\varepsilon_{crit}$ increases. Notably, ablated networks with path redirection are qualitatively more similar to the original fiber network when compared with ablated fiber network with a new shortest path. This basic study on the sensitivity of the mechanics of random fiber networks to fiber ablations on shortest paths opens up a new potential tool to understand remodeling in biological networks and logical design of architected materials.}

\section{Conclusion}
\label{sec:conclusion}

In this work, we mapped the structure-function relationships of random fiber networks by connecting local heterogeneous geometric features to mechanical function. To accomplish this goal, we implemented a finite element analysis based computational framework that can handle numerical convergence difficulties encountered in random fiber network simulations, such as shear locking, extrapolation locking, and local buckling in open source finite element analysis software FEniCS \citep{alnaes2015fenics,logg2012automated}. With our robust computational model of random fiber networks in hand, we then associate the heterogeneous geometric structure of random fiber networks to emergent mechanical function under loading. Initially, we investigated the structure-function relationships of single fiber chains, the building blocks of random fiber networks. We then link the geometric features of single fiber chains to their mechanics through a simple analytical reduced order model that combines an axial spring and a rotational spring. Additionally, we show that the geometric features of single fiber chains can predict their critical strain-stiffening transition point. Next, we translate our findings in single fiber chains to random fiber networks by extracting the distance-weighted shortest paths of random fiber networks and analyze them as single fiber chains in parallel. We are then able to combine our algorithm for identifying shortest paths with our analytical reduced order model to predict the mechanical response of random fiber networks. Additionally, we demonstrated that mechanically the shortest paths behave as force chains, and recruit sequentially to contribute to the strain-stiffening effect in random fiber networks. The results from this work demonstrate the connection between the geometric structure of random fiber networks and mechanical function via distance-weighted graph shortest paths.

Though our work is an important step towards mapping the structure-function relationships of random fiber networks, there are several key related areas that remain unexplored. \changes{For example, we only analyzed the structure-function relationships of Voronoi networks, but many other types of networks exist.} In general, for biological systems, the main types of fiber networks are fibrous networks and cellular networks, and both types of networks exhibit different mechanical behavior under loading \citep{heussinger2007role,islam2018effect}. In addition, many fiber networks, especially in biological systems, are embedded within elastic matrices \citep{kakaletsis2023mechanics}, and more work is needed to determine if the structure-function relationships found in this work will hold in the context of embedded fiber networks. And, most crucially, we only investigated structure-function relationships in the case of uniaxial tension -- a straightforward boundary condition. However, biological networks are subjected to complex boundary conditions, and extending our work to a larger set of boundary conditions is critical when applying the results from our work to study biological systems.

Despite the limitations of our work, the results presented here are a starting point for multiple future directions. For example, minor changes to the geometric structure such as changes to fiber orientation are known to affect mechanical response~\citep{hatami2009effect}. Thus, understanding how fiber orientation and re-orientation affect structure-function relationships is crucial when investigating biological process such as remodeling of fibrous tissue \citep{gizzi2024evolution, kuhl2005remodeling}, wound healing \citep{das2021extracellular}, and cell signaling \citep{beroz2017physical, berthier2024nonlinear, humphries2017mechanical, mann2019force}. Building on recent work that has experimentally observed changes in fiber orientation measured via distribution functions \citep{das2021extracellular, rubbens2009quantification}, our work can complement existing methods by directly connecting \emph{local} changes in structure and orientation to function through emerging load paths. Namely, extensions of this work could lead to understanding how structural perturbations from either fiber remodeling or external damage (i.e., wounding) change load paths and thus enable mechanical signaling through fiber network mechanics. Broadly speaking, this work explores the mechanical behavior of mesoscale structures and the relationship between mechanical behavior and complex local geometric features. Looking forward, we envision that understanding these systems is not only relevant to understanding the role of geometry and mechanics in biological processes that rely on long range force transmission and signaling, but also to designing architected materials that mimic these functionalities. In particular, designing mesoscale architected materials exhibiting ``mechanical intelligence'' \citep{jiao2023mechanical,lejeune2023locality, sitti2021physical} that can recapitulate the mechanical signaling behavior of living systems is a compelling area of future research. To this end, we have made our code available on Github (\url{https://github.com/pprachas/struc_func_fiber_network}) under an open-source license such that other researchers can build directly on our work to address these interesting adjacent problems.

\section{Declaration of competing interest}
The authors declare that they have no known competing financial interests or personal relationships that could have appeared to influence the work reported in this paper.

\section{Acknowledgements}
This work was made possible with funding through the Boston University David R. Dalton Career Development Professorship, the Hariri Institute Junior Faculty Fellowship, the Haythornthwaite Foundation Research Initiation Grant, and the National Science Foundation Grant CMMI-2127864. This support is gratefully acknowledged. We also acknowledge the support of Boston University's Research Computing Services for providing computing resources.

\appendix

\section{\changes{Shear-Deformable Geometrically Exact Beams in 3D}}
\label{appendix:meth_beams}
\begin{figure}[h]
    \centering
    \includegraphics[width= 0.95\textwidth]{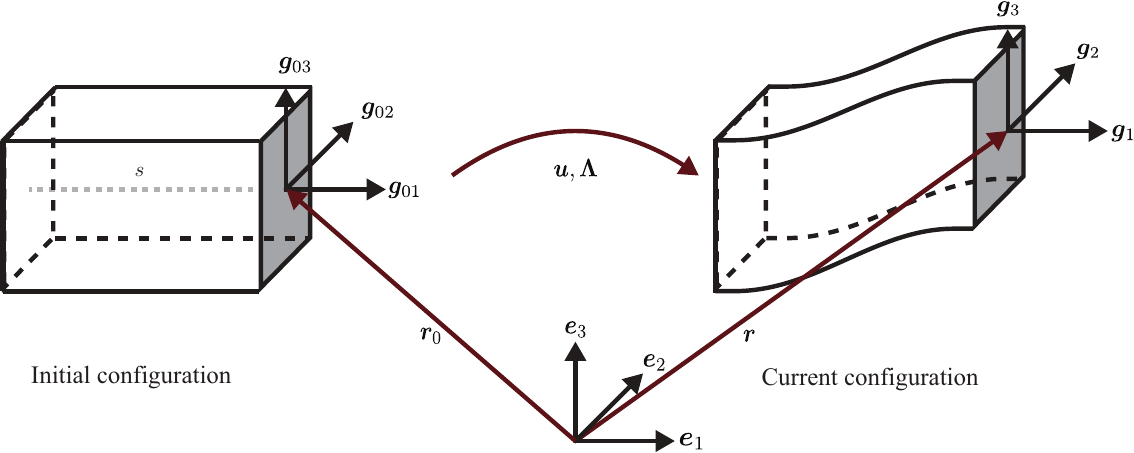}
    \caption{\changes{Schematic of the kinematics defining initial configuration and deformed configuration according to geometrically exact beam theory. The initial beam position $\bm{r}_0(s)$ is mapped to the current beam configuration $\bm{r}(s)$ through the displacement field $\bm{u}$, and the initial beam orientation $\{ \bm{g}_{01}, \bm{g}_{02}, \bm{g}_{03} \}$ is mapped to the current beam orientation $\{ \bm{g}_1, \bm{g}_2, \bm{g}_3 \}$ through the rotation field $\bm{\Lambda}$.}}
    \label{fig:meth_beams}
\end{figure}

\changes{We model our single fiber chains and random fiber networks using shear-deformable geometrically exact beams. In brief, starting from kinematics, the position of Simo-Reissner beams can be described with a parameterized beam centerline curve in the undeformed beam configuration $s \in [0,\ell]$, where $\ell$ is the beam length, and the beam orientation is defined through a field of right-handed orthonormal triads $\{ \bm{g}_1(s), \bm{g}_2(s), \bm{g}_3(s) \} \in \mathbb{R}^3 $. The initial beam configuration with respect to the global initial coordinates is $\bm{r}_0(s) \in \mathbb{R}^3$, and the current beam configuration is described by $\bm r(s) \in \mathbb{R}^3$ (see Fig. \ref{fig:meth_beams}). The initial beam configuration can be mapped to the current beam configuration through the displacement field $\bm{u} \in \mathbb{R}^3$ such that:}
\changes{
\begin{equation}
 \bm{r}(s) = \bm{r}_0(s) + \bm{u}(s) \, .
\end{equation}}
\changes{
Typically, the direction of the first initial triad is the tangent direction of the beam (i.e., $\bm{g_{01}} = \frac{d\bm{r}_0}{ds} = \bm{r}_{0,s}$) and the other two triads are the principal axes of the beam. The rotation tensor field $\bm{\Lambda} \in SO(3)$ maps the rotation from the initial orthonormal triads (i.e., the material triads) $\{\bm{g}_{01}, \bm{g}_{02},  \bm{g}_{03} \}$ to the current orthonormal triads (i.e., the spatial triads) $\{\bm{g}_1, \bm{g}_2,  \bm{g}_3 \}$ such that:}

\changes{
\begin{equation}
    \bm{g}_{i} = \bm{\Lambda} \bm{g}_{0i} \quad ; \quad i = 1,2 \, .
\end{equation}}
\changes{
The strain measures are defined as:}
\changes{
\begin{equation}
\begin{aligned}
    \bm{\varepsilon} &= \bm{\Lambda}^T \bm{r}_{,s} - \bm{g}_{01} \\
    \bm{\chi} &= \text{axial} \left( \bm{\Lambda}^T  \bm{\Lambda}_{,s} \right)
\end{aligned}
\end{equation}}
\changes{
where $\bm{\varepsilon}$ and $\bm{\chi}$ are the translational and rotational strain respectively, while $\text{axial}(.)$ denotes the axial vector $\bm{a} = [a_1,a_2,a_3]^T$ such that for a skew symmetric tensor $\bm{A}$:}

\changes{
\begin{equation}
    \bm{A} = \begin{bmatrix}
    0 & -a_3 & a_2\\
    a_3 & 0 & -a_1\\
    -a_2 & a_1 & 0
    \end{bmatrix} \, .
\end{equation}}

\changes{
We note that for any $\bm{\Lambda} \in SO(3)$, $\bm{\Lambda}^T \bm{\Lambda}_{,s}$ is skew symmetric, and the axial vector corresponds the rotational degrees of freedom in the beam. Additonally, we also note that in the general case of 3D geometrically exact beams, dealing with 3D rotations and rotation vector parameterization can lead to singularities, lack of objectivity, and path dependence. Interested readers should refer other work on dealing with rotation parameterization, which is outside the scope of this work \citep{ibrahimbegovic1997choice,magisano2020large, meier2019geometrically}. Specific implementation details and code for rotation parameterization in FEniCS can be found in \citep{bleyer2018numericaltours}, and its integration with numerical continuation methods can be found in \citep{Prachaseree_FEniCS-arclength_A_numerical_2024}. As proposed by Simo \citep{simo1985finite}, the constitutive matrices of the beams are:}

\changes{
\begin{equation}
\begin{aligned}
    \bm{C}_N &= \begin{bmatrix}
    EA & 0 & 0 \\
    0 & \mu A_2^* & 0 \\
    0 & 0 & \mu A_3^*
    \end{bmatrix}\\
    \bm{C}_M &= \begin{bmatrix}
    \mu J  & 0 & 0 \\
    0 & \;EI_2\; & 0 \\
    0 & 0 & EI_3
    \end{bmatrix}
\end{aligned}
\end{equation}}

\changes{
where $E$ and $\mu$ are the Young's Modulus and shear modulus respectively, $A$ is the cross-section area, $A^*_2$ and $A^*_3$ are the effective shear-corrected areas, $J$ is the polar area moment of inertia, and $I$ is the second area moment of inertia.
The constitutive law is a hyperelastic strain energy density function expressed as:}

\changes{
\begin{equation}
\tilde{\Pi}_{int} = \frac{1}{2} \biggl( \bm{\varepsilon} \bm{\cdot} \bm{C}_N \bm{\varepsilon} + \bm{\chi} \bm{\cdot} \bm{C}_M \bm{\chi} \biggr) \, .
\end{equation}}

\changes{
We note that even though the material constitutive law is linear elastic there are still geometric nonlinearities present in the system due to finite rotations.
The beam virtual work equation can then be written as $\delta \bigl[ \int_l (\tilde{\Pi}_{int}   - \tilde{\Pi}_{ext} ) \; ds \bigr]  = 0$ , where $\delta$ denotes taking the variation and $\tilde{\Pi}_{ext}$ denotes the external forces. Note the $\delta \tilde{\Pi}_{ext}$ can be non-trivial due to the presence of rotations and moments. In general, due to the non-conservative nature of applied moments, the tangent stiffness matrix is not guaranteed to be symmetric. Interested readers should refer to \citep{meier2019geometrically} for treatment of these terms. In our case, $\delta \Pi_{ext} = 0$ since we apply incremental displacements as our form of external loading. as such, for brevity and without loss of generality, all subsequent derivations and analysis will ignore traction boundary conditions.}
\changes{
Finally, the stress resultants are:}

\changes{
\begin{equation}
\begin{aligned}
    \bm{N} &= \frac{\partial \tilde{\Pi}_{int}}{\partial \bm{\varepsilon}} = \bm{C}_N \bm{\varepsilon} \\
    \bm{M} &= \frac {\partial \tilde{\Pi}_{int}}{\partial \bm{\chi}}  = \bm{C}_M \bm{\chi}
\end{aligned}
\end{equation}}

\changes{
where $\bm{N}$ captures forces, and $\bm{M}$ captures moments.}

\section{Additional Details on Tikhonov Regularization}
\label{appendix:regularization}

\begin{figure}[h]
    \centering
    \includegraphics[width= \textwidth]{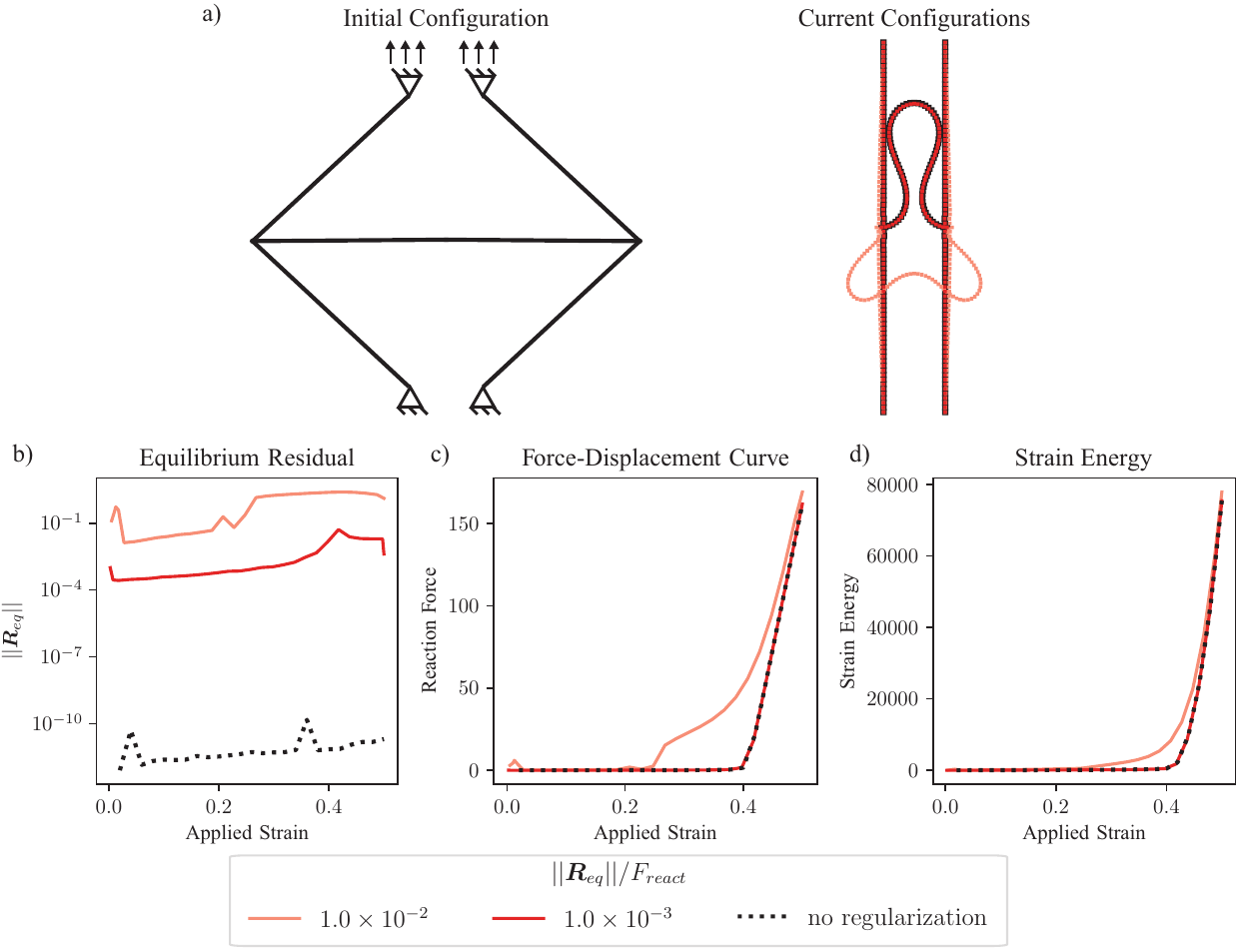}
    \caption{\changes{Effect of regularization on a simple representative example. The regularized solutions (pink and red) are compared with the solution with no regularization (dotted black line). As an illustrative example, the regularization term for the pink line is tuned to be over-regularized to give a different solution than the equilibrium solution. a) Equilibrium residual $||\bm{R}_{eq}||$ vs. applied strain for different convergence tolerance; b) Force vs. applied displacement for different convergence tolerance; c) Strain Energy vs. applied displacement for different convergence tolerance d) Comparison of deformed configurations for both regularized and not regularization solutions. }}
    \label{fig:sensitivity_tri}
\end{figure}

\begin{figure}[h]
    \centering
    \includegraphics[width= \textwidth]{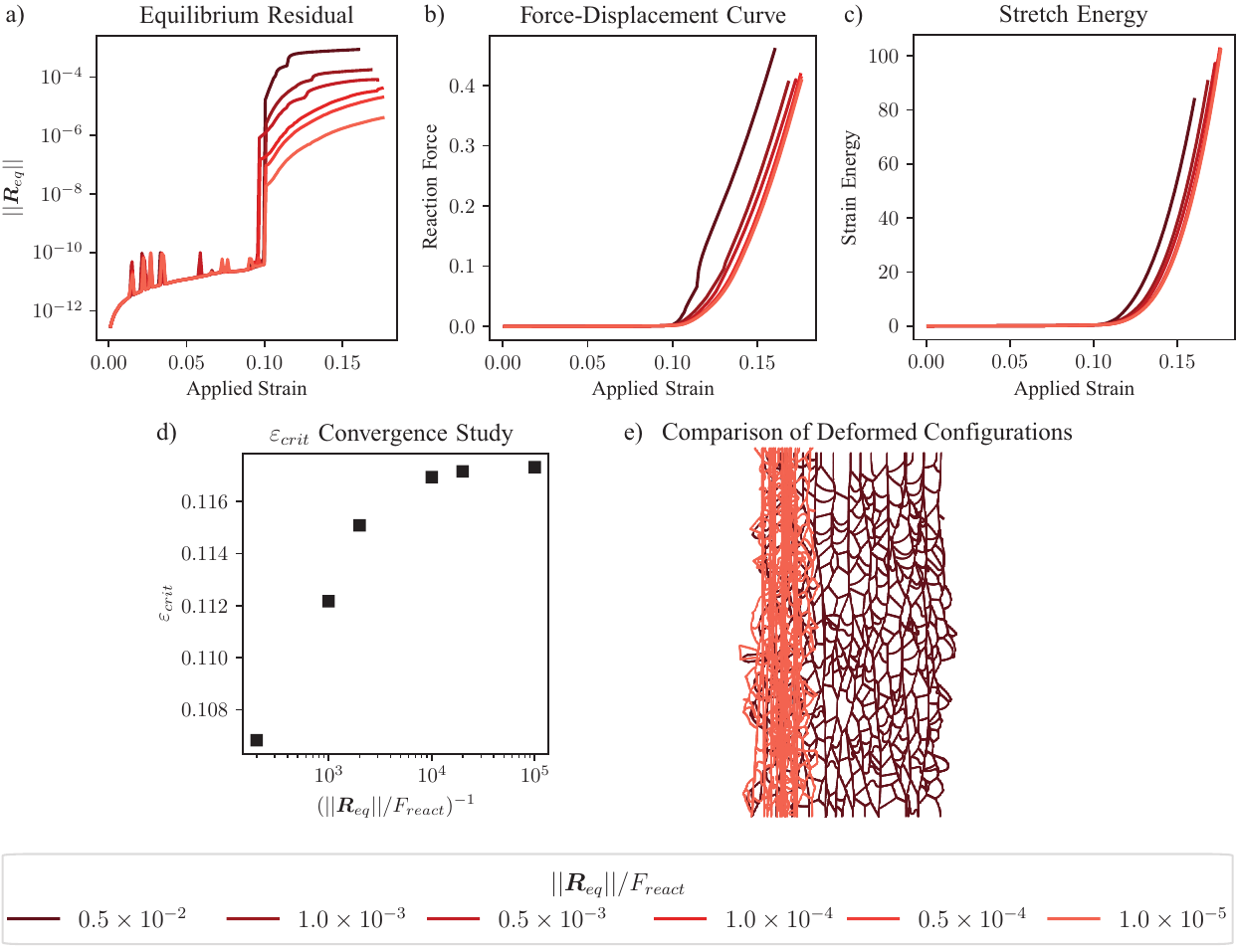}
    \caption{Sensitivity analysis for the convergence tolerance $||\bm{R}_{eq}||/F_{reac}$ in Tikhonov regularization on a representative random fiber network: a) Equilibrium residual $||\bm{R}_{eq}||$ vs. applied strain for different convergence tolerance; b) Force vs. applied displacement for different convergence tolerance; c) Strain Energy vs. applied displacement for different convergence tolerance; d) Critical transition strain $\varepsilon_{crit}$ for difference convergence tolerance; e) Deformed Configuration for the same random fiber networks with $||\bm{R}_{eq}||/F_{reac} = 0.5 \times 10^{-2}$ (dark red) and $||\bm{R}_{eq}||/F_{reac} = 1.0 \times 10^{-5}$ (pink). }
    \label{fig:sensitivity_tik}
\end{figure}

In Section \ref{sec:meth_fea}, we showed that $c_0$ controls the strength of regularization in our FEA model. In brief, from Eqn.~\ref{eqn:static_condense}, we show that Tikhonov regularization facilitates numerical convergence by ensuring that the equivalent stiffness matrix $\bm{K}$ is positive definite with a strong enough regularization (Eqn. \ref{eqn:static_condense}). However, a regularization term that is too large will fundamentally change the finite element problem that we are solving because it alters both the equivalent stiffness matrix $\bm{K}$ and the residual (see Eqn. \ref{eqn:discretized_stationary}). As such, choosing an appropriate value of $c_0$ is a delicate balance between promoting numerical convergence and corrupting the original finite element problem. To control the strength of regularization, we initially set $c_0 = 10^{-7}$ and increase/decrease $c_0$ based on the convergence of the equilibrium residual $||\bm{R}_{eq}||/F_{react}$. Specifically, if we lower the tolerance $||\bm{R}_{eq}||/F_{react}$, then the regularization strength will also be decreased to accommodate for this stricter convergence criterion.

\changes{As a start, we investigate the effect of the regularization term on a simple representative structure with similar deformation behavior as random fiber networks as shown in Fig. \ref{fig:sensitivity_tri}a. Intuitively, as uniaxial tension is applied to the structure, the beams on the outer edge will straighten, compressing the horizontal beam, making the horizontal beam buckle in a similar fashion to a local instability in random fiber networks. Note that to induce buckling we break the symmetry of the horizontal beam by geometrically perturbing the central node of the structure in the initial configuration. For this simple representative example, we examined the effect of regularization for $||\bm{R}_{eq}||/F_{react}\leq \{ 10^{-2}, 10^{-3} \}$. Additionally, since this simple problem can converge without any regularization, we use the unregularized solution as the ground truth. Fig. \ref{fig:sensitivity_tri}b verifies our regularization implementation that a smaller $||\bm{R}_{eq}||/F_{react}$ results in a smaller equilibrium residual, with the smallest equilibrium residual belonging to the unregularized solution as expected. With our implementation verified, we now look at the effect of the convergence criterion $||\bm{R}_{eq}||/F_{react}$ on the force vs. strain curve (Fig. \ref{fig:sensitivity_tri}c) and strain energy vs. strain curve (Fig. \ref{fig:sensitivity_tri}d). Qualitatively, both the force vs. strain and strain energy vs. strain curves match the regularized solution relatively well for $||\bm{R}_{eq}||/F_{react} = 10^{-3}$ (i.e., weaker regularization), but the force vs. strain and the strain energy vs. strain curve from $||\bm{R}_{eq}||/F_{react} = 10^{-2}$ (i.e., stronger regularization) does not match the unregularized solution. We also compared the deformed configuration of all three cases in Fig \ref{fig:sensitivity_tri}a and similarly, the solution from $||\bm{R}_{eq}||/F_{react} = 10^{-3}$ matches the unregularized solution well but $||\bm{R}_{eq}||/F_{react} = 10^{-2}$ does not match the unregularized solution and demonstrates the effect of over-regularizing the solution. From Eqn. \ref{eqn:regularization} the Tikhonov regularization term penalizes solutions that are far away from the previously converged solution, which in this simple case of local instability (i.e., which entails large deformations) would encourage the buckling beam to not deform as much, resulting in a buckling mode shape that may not be realized physically. This simple example also demonstrates that a small enough regularization will not affect the overall solution.}

Next, to understand how the additional Tikhonov regularization term affects our FEA solution of random fiber networks, we perform sensitivity analysis of the additional convergence criterion $||\bm{R}_{eq}||/F_{react}\leq \{0.5\times10^{-2},1.0\times10^{-3},0.5\times10^{-3},1.0\times10^{-4}, 0.5\times10^{-4}, 1.0\times10^{-5} \}$ on the FEA solution. \changes{Since we demonstrated from the previous simple example that a larger regularization term affects the solution, the regularization term is only applied to the problem when the Newton solver fails to converge for a small applied displacement step. Note that in this case, the random fiber network does not converge at the maximum applied strain without a regularization term, resulting in no ground truth to act as comparison. However, as demonstrated in the previous example, a small enough regularization term corresponds to the unregularized solution.} As a representative example to show for this sensitivity study, we use a random fiber network with $n=500$ initial Voronoi seeds (i.e., the densest fiber networks studied in this work). First, to verify our implementation, we examine $||\bm{R}_{eq}||$ with decreasing convergence criterion $||\bm{R}_{eq}||/F_{react}$ in Fig. \ref{fig:sensitivity_tik}a. As expected, a smaller convergence criterion leads to a smaller magnitude of the equilibrium residual $||\bm{R}_{eq}||$ (note that the y-axis in this plot is logarithmic). \changes{Additionally, the equilibrium residual is initially low before experiencing a sharp increase after some applied strain, which confirms that the regularization is used only once the Newton solver can no longer converge.} With our implementation verified, we now look at the effect of the convergence criterion $||\bm{R}_{eq}||/F_{react}$ on the force vs. strain curve (Fig. \ref{fig:sensitivity_tik}b) and strain energy vs. strain curve (Fig. \ref{fig:sensitivity_tik}c). In general, a larger convergence criterion leads to larger force and strain energy values, and both the force vs. displacement and strain energy vs. displacement curves converge for smaller values of $||\bm{R}_{eq}||/F_{react}$. 

Additionally, we look at the convergence of the critical strain transition point $\varepsilon_{crit}$ with respect to $\left(||\bm{R}_{eq}||/F_{react} \right)^{-1}$ in Fig. \ref{fig:sensitivity_tik}d and observe that $\varepsilon_{crit}$ also converges with smaller $||\bm{R}_{eq}||/F_{react}$. Finally, to qualitatively understand the effect of the regularization on the FEA solution, we visualize the deformed configuration for the solutions with $||\bm{R}_{eq}||/F_{react} \leq 0.5\times10^{-2}$ (i.e., stronger regularization) and $||\bm{R}_{eq}||/F_{react} \leq 1.0\times10^{-5}$ (i.e., weaker regularization). The deformed configurations shows that the fiber network with stronger regularization will have a smaller Poisson's effect. Overall, this study shows that choosing an appropriate regularization parameter $c_0$ is a delicate balance between numerical convergence and altering the finite element problem. We concluded that using $||\bm{R}_{eq}||/F_{react} \leq 0.5\times10^{-4}$ is sufficient and appropriate for this work. 

\section{Mesh Refinement Study}
\label{appendix:mesh_refinement}

\begin{figure}[p]
    \centering
    \includegraphics[width= \textwidth]{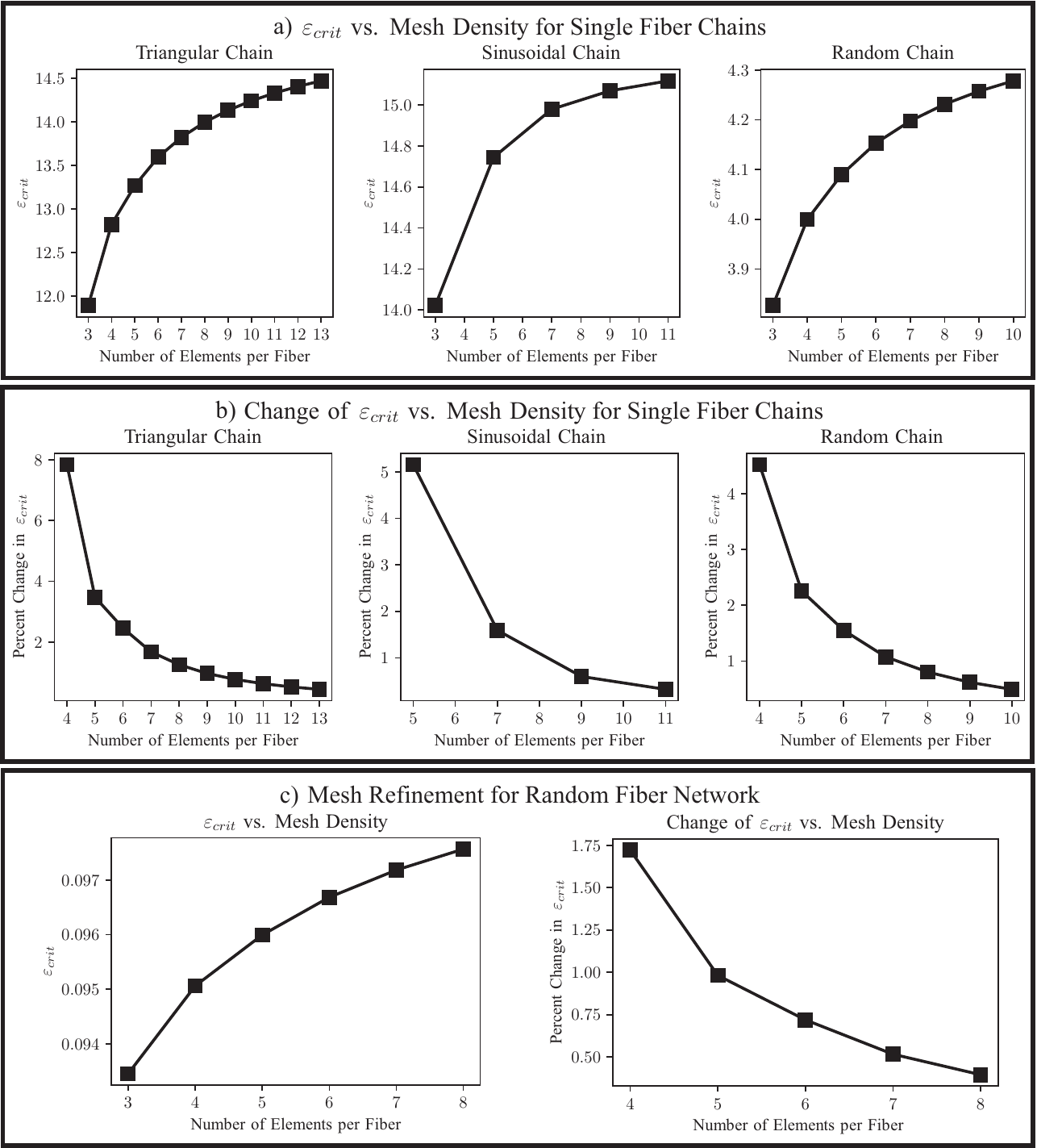}
    \caption{Mesh refinement studies on systems studied in this work : a)  Critical strain transition  $\varepsilon_{crit}$ with different mesh densities for representative single fiber chains all types of single fiber chains used in our work; b) Percent change of $\varepsilon_{crit}$ as we increase the mesh density for all $3$ types of single fiber chains. c) Mesh refinement study for representative random fiber network. Left plot visualizes critical strain transition  $\varepsilon_{crit}$ vs. different mesh densities, and the right plot visualizes Percent change of $\varepsilon_{crit}$ vs. mesh density  }
    \label{fig:mesh_refinement}
\end{figure}

In this work, we investigate the mechanics of single fiber chains and random fiber networks. To make sure our the FEA results are agnostic to mesh density, we perform mesh refinement studies on both systems. Starting with single fiber chains, to highlight that our mesh refinement study covers all single fiber chains analyzed in this work, we select an example from each chain type to show the results of mesh refinement. To make sure that our mesh refinement study encompasses all single fiber chains, we chose representative examples with the most complex geometries, and thus the most difficult to converge numerically. Specifically, in Fig. \ref{fig:mesh_refinement}, we show the results of a mesh refinement study for chains with parameters $\lambda = L_0/40$, $a = L_0/10$ in the triangular and sinusoidal cases, and $n_c = 80, w = L_0/10$ for the random case. We also set $\tilde{\kappa} = 10^{-6}$ for all chains because that corresponds to the most slender structures and the condition number of the stiffness matrix scales with beam slenderness \citep{meier2019geometrically}, with more slender beams having a higher condition number (i.e., low numerical stability). Following previous work \citep{parvez2024methodological}, the fiber chains are discretized by splitting each fiber into elements of uniform length. Note that in this case the number of \textit{elements} will be the same for all fibers in the chain, but the length of each element varies in the chain. To perform mesh refinement, we introduce an additional node to the fiber and ensure uniform element length (i.e., h-refinement). In the case of sinusoidal chains, one ``fiber'' is half the wavelength of the sinusoid, and we also add $2$ nodes (and subsequently $2$ elements) per refinement step since we are approximating a curved geometry with straight Lagrange elements (i.e., we are also discretizing the geometry alongside the solution). 
We use $\varepsilon_{crit}$ as our quantity of interest and define the percent change in $\varepsilon_{crit}$ as:

\begin{equation}
\frac{|\varepsilon_{crit}^{current}-\varepsilon_{crit}^{prev}|}{\varepsilon_{crit}^{prev}}\times 100 \% \, 
\end{equation}

where $\varepsilon_{crit}^{prev}$ and $\varepsilon_{crit}^{current}$ are the previous and current values of $\varepsilon_{crit}$ respectively.
We terminate the mesh refinement study when the percent change of $\varepsilon_{crit}$ per step less than $0.5 \%$. Fig. \ref{fig:mesh_refinement} shows that $13$ elements per fiber for triangular chain, $11$ elements per fiber for sinusoidal chains, and $10$ elements per fiber for random chains is sufficient for a converged solution following this mesh refinement criteria.

For random fiber networks, we again select a representative example that is the most difficult to numerically converge and has sufficiently complex geometry (i.e., the most dense network). As such, with similar arguments to the case of single fiber chains, we chose a $n=500$ fiber network with $\tilde{\kappa}  = 10^{-6}$. We discretized our fiber networks in a similar manner to single fiber chains where we split each fiber into a prescribed number of elements with uniform length, and h-refinement is performed by sequentially increasing the number of nodes in each fiber. Similarly, we use $\varepsilon_{crit}$ as our quantity of interest and terminate our study when the percent change of $\varepsilon_{crit}$ is less than $0.5\%$. From Fig. \ref{fig:mesh_refinement}c, we conclude that $8$ elements per fiber is sufficient for a converged solution. However, while $8$ elements per fiber is sufficient for solution convergence, we found that this mesh density can have trouble converging for other random fiber network samples due to local instabilities in random fiber networks. As a result, we used a higher mesh density to help deal with the local instabilities, and empirically chose $15$ elements per fiber as our mesh density for our numerical experiments on random fiber networks. 

\section{Additional Error Analysis of Analytical Model for Single Fiber Chains}
\label{appendix:single_abc}

\begin{figure}[h]
    \centering
    \includegraphics[width= \textwidth]{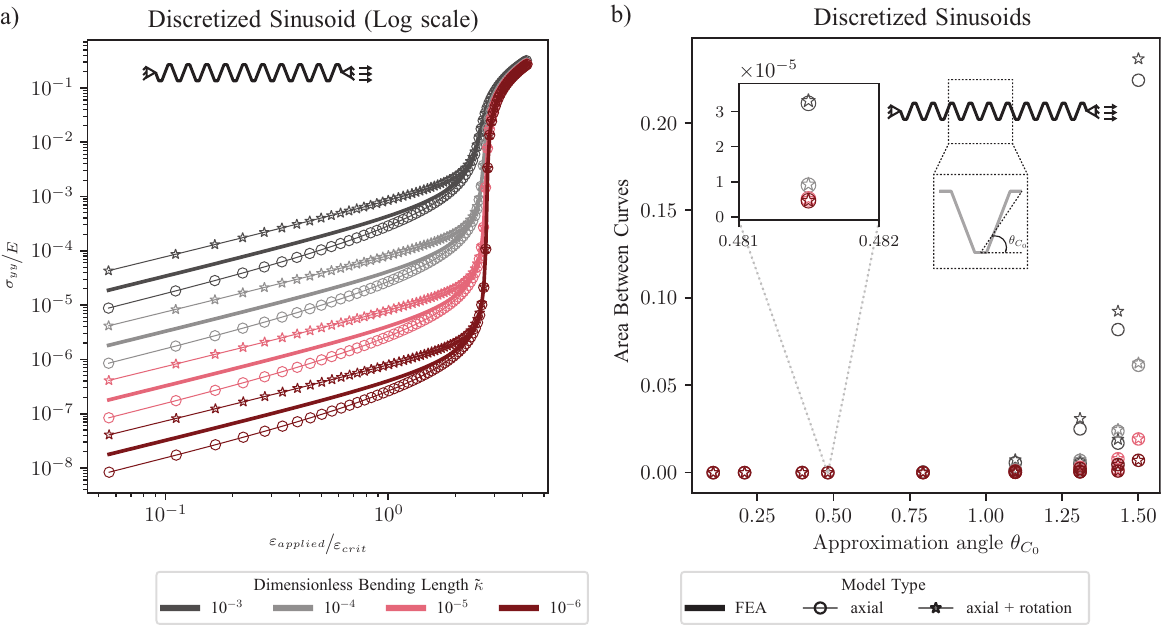}
    \caption{Additional analysis of our analytical reduced order model through a special example of a discretized sinusoidal chain: a) Visualizes the normalized force $\sigma_{yy}/E$ vs.strain $\varepsilon_{applied}/\varepsilon_{crit}$ from the FEA model and reduced order model; b) Absolute area between normalized force vs. strain curves with different approximation angle $\theta_{C_0}$}
    \label{fig:appendix_single_error}
\end{figure}

In Section \ref{sec:res_analytical_single}, we discovered that the full analytical reduced order model that combines an axial spring and a rotational spring matches the FEA results better than the partial reduced order model that only consists of the axial spring in most cases. However, in certain rare cases for random fiber chains, the additional rotational spring term is detrimental to the accuracy of the reduced order model. As a representative simple example where the rotational spring term does not improve the reduced order model, we look at a single fiber chain formed by discretizing a sinusoidal wave (see Fig. \ref{fig:appendix_single_error}a inset). More specifically, this chain is formed by sampling $3$ points per wavelength from a sinusoid that are uniformly spaced in the same direction of the applied load. This will form a chain similar to a triangular chain but with an additional small straight fiber section that is in the same direction as the loading direction. The normalized force vs. displacement plot of the structure is visualized in Fig. \ref{fig:appendix_single_error}a. In general, the partial reduced order model underestimates the FEA result while full reduced order model overestimates the results. We generalized this notion by looking at different samples of discretized sinusoidal chains with $\lambda = \{ L_0, L_0/5,L_0/10,L_0/20,L_0/40\}$ and $a = \{ L_0/10, L_0/20, L_0/40 \}$ (see Eqn. \ref{eqn:generate_periodic}). Again, the trend of the additional rotational term overestimating the normalized force $\sigma_{yy}/E$ (i.e., higher absolute area between curves) is observed. 

A simple explanation for this overestimation can be deduced from the geometry of the discretized sinusoidal chains. First, from both Fig. \ref{fig:res_single_mech}c and Fig. \ref{fig:appendix_single_error}, the discrepancy between the analytical reduced-order model and the FEA result is in the bending dominating phase. Physically, in the bending dominated phase, the straight section of the discretized sinusoidal chain will not rotate (and will not contribute much to the reaction force) due to symmetries in the structure. As such, the bending force from the discretized sinusoidal will be similar even if we remove the straight section of fibers, which will effectively turn the chain from a discretized sinusoidal chain into a triangular chain. Note that a for a triangular chain $\ell_c$ and $\theta_C$ are computed exactly by definition. However, because the straight sections are short, they reduce $\ell_c$, which in turn \emph{increases} the bending stiffness when compared to the equivalent triangular chain, thus leading to overestimation of the bending force. 

\section{Additional Error Analysis of the Analytical Model for Random Fiber Network}
\label{appendix:fiber_abc}

\begin{figure}[h]
    \centering
    \includegraphics[width= \textwidth]{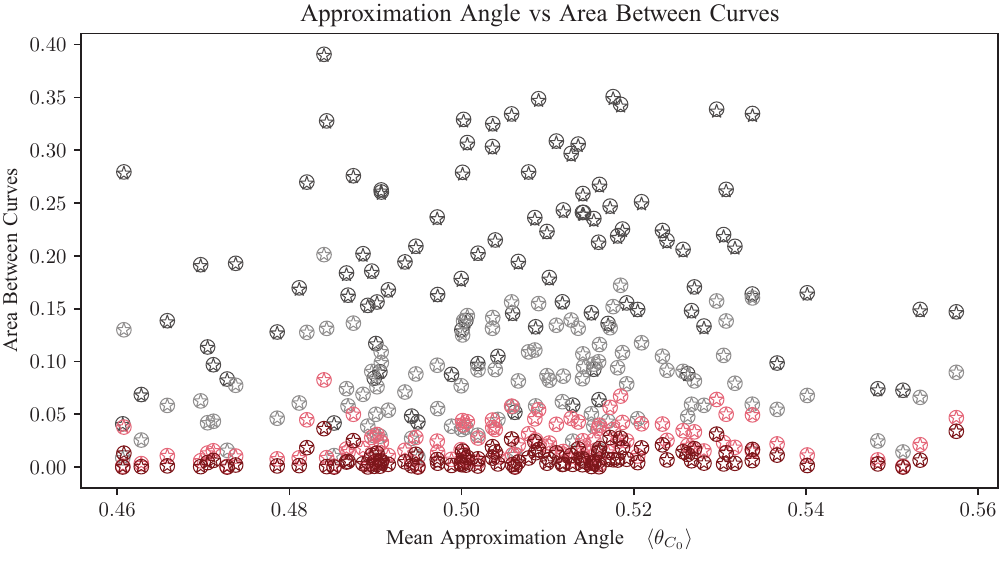}
    \caption{Absolute area between the FEA result and analytical reduced order model with different mean approximation angle $\langle \theta_{C_0} \rangle$}
    \label{fig:app_fiber_abc}
\end{figure}

In addition to analyzing the absolute area between curves vs. domain size $L_0/\ell_c$ in Section \ref{sec:res_fiber_abc}, we also look at the absolute area between curves vs. mean approximation angle of the shortest paths $\langle \theta_{C_0} \rangle$, where the mean approximation angle for $k$ shortest paths is defined as:

\begin{equation}
\langle \theta_{C_0} \rangle = \frac{1}{k} \sum_{k} \arccos \left( \frac{L_{0}^k}{L_{C_0}^k} \right) \, .
\end{equation}

In Fig. \ref{fig:app_fiber_abc}, we plot absolute area between curves vs. mean approximation angle of the shortest paths $\langle \theta_{C_0} \rangle$ and found that there is no clear correlation between the absolute area between curves and $\langle \theta_{C_0} \rangle$. As such, contrary to the singe fiber chains case (i.e., Fig \ref{fig:res_MAPE}), the waviness of the shortest paths do not contribute the discrepancy between our reduced order model and FEA result. 

\section{Additional $\tilde{\kappa}$ vs. $\varepsilon_{crit}$ for Fiber Networks}
\label{appendix:fiber_phase}

\begin{figure}[h]
    \centering
    \includegraphics[width= \textwidth]{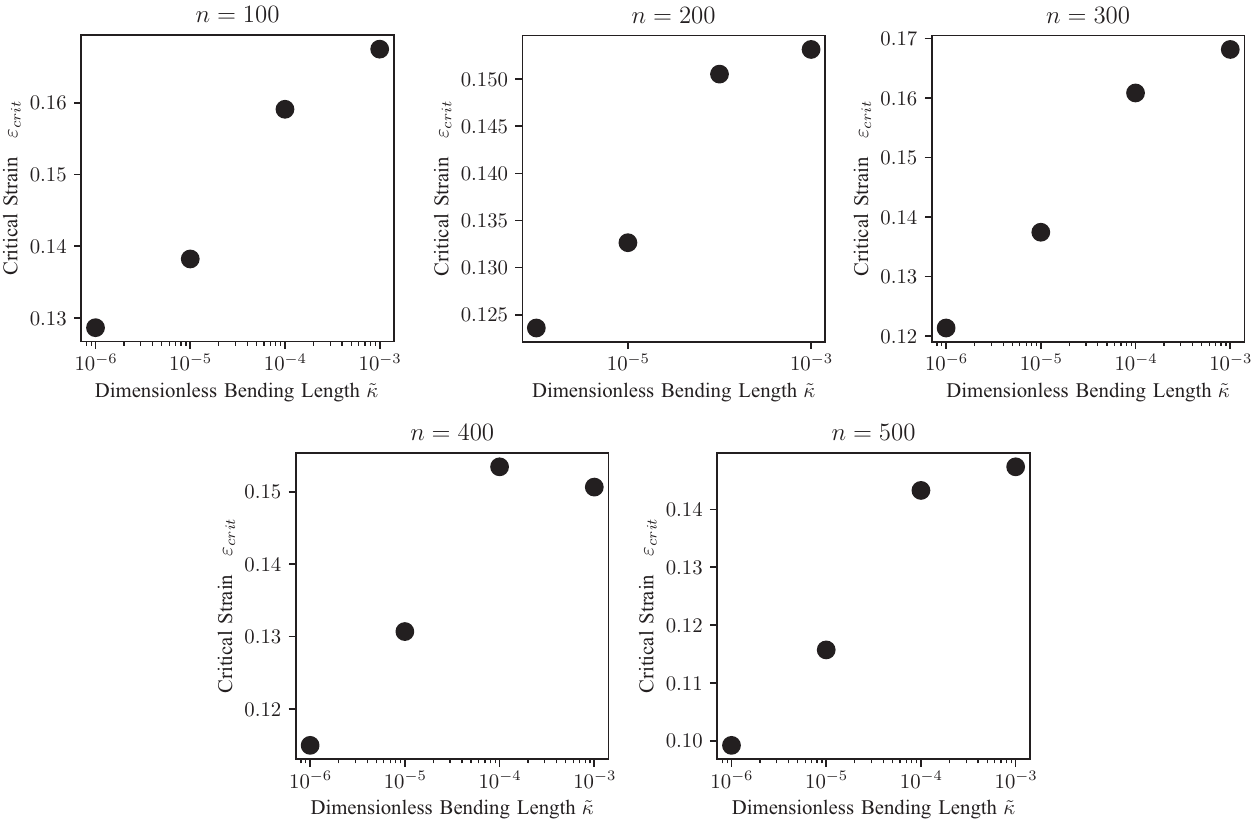}
    \caption{Relationship of critical strain $\varepsilon_{crit}$ and geometric parameter $\tilde{\kappa}$ for a representative random fiber networks with $n \in \{ 100,200,300,400,500 \}$. The plots for each $n$ are from the same random fiber network geometry.}
    \label{fig:app_phase}
\end{figure}

To further support the notion that a phase diagram analogous to Fig. \ref{fig:res_single_mech} is non-trivial for the case of random fiber networks, we look at the dimensionless bending length $\tilde{\kappa}$ vs. critical strain $\varepsilon_{crit}$ for representative fiber networks with $n= \{ 100,200,300,400,500 \}$ in Fig. \ref{fig:app_phase}. Note for each $n$ the same fiber network geometry is used. From the plots, there is a clear monotonically increasing relationship between $\tilde{\kappa}$ and $\varepsilon_{crit}$ for most cases except for the representative case $n=400$. As a result, the complicated relationship between  $\tilde{\kappa}$, $\tau_0^k$ and $\varepsilon_{crit}$ makes it difficult to construct a phase diagram that determines the dominant energy mode from geometric parameters. Specifically, the complicated relationship between  $\tilde{\kappa}$, $\tau_0^k$ and $\varepsilon_{crit}$ for random fiber networks means that there are not clear phase boundaries similar to the case of single fiber chains (see Fig. \ref{fig:res_phase} for phase diagram of single fiber chains).

\newpage
\bibliographystyle{plain}
\bibliography{main}

\end{document}